\documentclass{aptpub}
\usepackage[T1]{fontenc}
\usepackage{color}
\usepackage{amsfonts}
\usepackage{graphicx}
\usepackage{latexsym,color}
\authornames{
Valentin Patilea and Hamdi Ra\"{i}ssi }
\shorttitle{Causality tests with heteroscedastic errors}

\numberwithin{equation}{section}  

\numberwithin{figure}{section}
\begin{document}

\begin{center}
\title{
Adaptive estimation of vector autoregressive models with time-varying variance: application to testing linear causality in mean}
\end{center} 

\qquad

\begin{center}
Valentin Patilea$^{a}$ and Hamdi Ra\"{i}ssi$^{b}$\footnote{ 20, avenue des buttes de Coësmes, CS 70839, F-35708 Rennes Cedex 7,
France. Email: valentin.patilea@insa-rennes.fr and hamdi.raissi@insa-rennes.fr} 
\end{center}

\begin{center}
$^a$ IRMAR-INSA \& CREST-Ensai\\
$^b$ IRMAR-INSA
\end{center}

\qquad

\begin{center}
First version January 2010\\
This version July 2010\\
\end{center}

\quad

\begin{abstract}
Linear Vector AutoRegressive (VAR) models where the innovations could be unconditionally heteroscedastic and serially dependent are considered. The volatility structure is deterministic and quite general, including breaks or trending variances as special cases. In this framework we propose Ordinary Least Squares (OLS), Generalized Least Squares (GLS) and Adaptive Least Squares (ALS) procedures.
The GLS estimator requires the knowledge of the time-varying variance structure while in the ALS approach the unknown variance
is estimated by kernel smoothing with the outer product of the OLS residuals vectors. Different bandwidths for the different cells
of the time-varying variance matrix are also allowed. We derive the asymptotic distribution of the proposed estimators for the VAR
model coefficients and compare their properties. In particular we show that the ALS estimator is asymptotically equivalent to the
infeasible GLS estimator. This asymptotic equivalence is obtained uniformly with respect to the bandwidth(s) in a given range and
hence justifies data-driven bandwidth rules. Using these results we build Wald tests for the linear Granger causality in mean which
are adapted to VAR processes driven by errors with a non stationary volatility. It is also shown that the commonly used standard Wald
test for the linear Granger causality in mean is potentially unreliable in our framework.
Monte Carlo experiments illustrate the use of the different estimation approaches for the analysis
of VAR models with stable innovations.
\\
\end{abstract}



\qquad

\section {Introduction}
\label{S1}

In the recent years the study of linear time series models in the context of unconditionally heteroscedastic innovations has become of increased interest. This interest may be explained by the fact that numerous applied works pointed out that unconditional volatility is a common feature in economic data. For instance Doyle and Faust (2005), Ramey and Vine (2006), McConnell and Perez-Quiros (2000), Blanchard and Simon (2001) among other references, pointed out a declining volatility for many economic data since the 1980s. Sensier and van Dijk (2004) found that 80\% of 214 U.S. macroeconomic time series they considered exhibit a break in volatility.

In the univariate time series case Busetti and Taylor (2003), Cavaliere (2004), Cavaliere and Taylor (2007) and Kim, Leybourne and Newbold (2002) among other references, considered the test of unit roots with non stationary volatility, while Sanso, Arago and Carrion (2004) proposed  tests to detect volatility breaks in the residuals.
Robinson (1987) and Hansen (1995) studied univariate linear models with a non stationary volatility.
Phillips and Xu (2005) investigated the Ordinary Least Squares (OLS) estimation of univariate stable autoregressive processes.
Xu and Phillips (2008) considered the same model and proposed an Adaptive Least Squares (ALS)
approach which are based on nonparametric estimation of the
volatility of the innovations using OLS residuals. The main conclusion of Xu and Phillips (2008) is
that the ALS estimating approach could be much more effective than the OLS
estimation. They also found that the asymptotic behavior of the ALS estimator does not dependent on the volatility structure.
Multivariate processes are often
used in econometric applications because they allow to study cross-correlations between variables. In the multivariate framework Boswijk and Zu (2007) and Cavaliere, Rahbek and Taylor (2007) studied cointegrated systems in presence of
non stationary volatility.

In this paper we study the inference in linear vector autoregressive (VAR) models with volatility changes and possibly serially dependent innovations. Three methods for estimating the VAR coefficients are investigated: OLS, infeasible Generalized Least Squares (GLS) based on the knowledge of the time-varying volatility structure, and ALS which is defined like the GLS but using a kernel estimate of the volatility structure. The kernel smoothing could be used with a single bandwidth for the whole volatility matrix or with different bandwidths for different cells. In some sense, we extend the approach of Phillips and Xu
(2005) and Xu and Phillips (2008) to the VAR framework. In particular, we see that in the multivariate case the asymptotic distribution of the GLS and ALS estimators is no longer free from the time-varying volatility structure. Moreover, our asymptotic results are uniform with respect to the bandwidth in a given range. This opens the door to data-driven choices of the smoothing parameter, for instance by cross-validation. Such uniformity results seems new even for the univariate case.

As an application of the new estimation methodology, we also consider the problem of test linear causality in mean. The linear causality in
mean, introduced by Granger (1969), is often used to investigate causal relations between subsets of variables. For
instance Sims (1972), Feige and
Pearce (1979) or Stock and Watson (1989) studied the
money-income causality relation. Bataa \textit{et al.} (2009) studied the links between the inflations of different countries by testing linear causality relations. This can be explained by the fact
that linear causality in mean can be easily tested  by considering tests of zero restrictions
on the parameters of VAR models. However, the existing test procedures for checking the linear causality in mean are based on the iid innovation assumption, while several empirical analysis contradict this setting. For instance, Bataa \textit{et al.} (2009) underlined the presence of volatility breaks in their data set. In this paper, we use our theoretical results on the OLS and ALS estimation to propose new Wald tests for linear causality in mean adapted to the framework of non-stationary volatility. The asymptotic chi-square distribution of the new Wald type statistic obtained from the ALS approach is derived uniformly with respect to the bandwidth(s).

The structure of the paper is as follows.  Section \ref{S2} outlines the heteroscedastic VAR model, introduces the assumptions and the definitions of OLS and GLS estimators. Section \ref{S3} contains the results on the asymptotic behavior of the OLS and the infeasible Generalized Least Squares estimators. We also propose an estimator for the asymptotic variance of the OLS estimator. The ALS estimator based on kernel smoothing of OLS residuals is proposed in Section \ref{S4} as a feasible asymptotically equivalent version of GLS estimator.
The asymptotic equivalence between ALS and GLS estimators is proved uniformly in the bandwidths involved in volatility estimation. To prove this equivalence we use, among other technical arguments, a recent version of a uniform CLT for martingale differences arrays obtained by Bae \emph{et al.} (2010), Bae and Choi (1999). A procedure for estimating the asymptotic variance of the ALS estimator is also provided.
The application of the new inference methodologies to the test of the linear Granger causality in mean in the presence of time-varying volatility is presented in Section \ref{S5}. The benefit from using our new Wald type test statistics and the failure of the classical Wald test designed for iid innovations is illustrated through an example. In section \ref{S6} the finite sample properties of the
different tests considered in this paper are studied by mean of Monte Carlo experiments. The better precision of the ALS estimator when compared to the OLS estimator is also highlighted. The proofs are relegated to the appendix.

The following notations will be used throughout in the paper. We
denote by $A\otimes B$ the Kronecker product of two matrices $A$ and
$B$, and $A\otimes A$ by $A^{\otimes2}$. The vector obtained by
stacking the columns of $A$ is denoted $\mbox{vec}(A)$.
The symbol $\Rightarrow$ denotes the convergence in distribution and we denote by $\stackrel{P}{\longrightarrow}$ the convergence in probability. We denote by $[u]$ the integer part of a real number $u$. The determinant of a square matrix $A$ is denoted by $\det A$.

\section {The model and least squares estimation of the parameters}
\label{S2}

Let us consider the observations $X_{-p+1},\dots,X_0,X_1,\dots,X_T$
 generated by the following VAR model
\begin{eqnarray}\label{VAR}
&&{X}_t={A}_{1}{X}_{t-1}+\dots+{A}_{p}{X}_{t-p}+u_t\\&&
u_t=H_t\epsilon_t,\nonumber
\end{eqnarray}
where the $X_t$'s are $d$-dimensional vectors. The stability condition on the matrices $A_i$,
$\det {A}(z)\neq 0$ for all $|z|\! \leq \! 1$ with
${A}(z) \!=\! I_d-\sum_{i=1}^{p}\!{A}_{i} z^i$ and $I_d$ denotes the $d\times d$ identity matrix, is assumed to be hold.
For a random variable $x$ we define $\parallel
x\parallel_r=(E\parallel x\parallel^r)^{1/r}$, where $\parallel
x\parallel$ denotes the Euclidean norm.  We also define
$\mathcal{F}_{t}$ as the $\sigma$-field
generated by $\{\epsilon_s:s\leq t\}$. The following assumption on
the $H_t$'s 
and the
process $(\epsilon_t)$ gives the framework of our paper.

\quad

\textbf{Assumption A1:} \quad (i) The $d\times d$ matrices $H_{t}$ are invertible and satisfy $H_{[Tr]}=G(r)$,
where the components of the matrix $G(r):=\{g_{kl}(r)\}$ are
measurable deter\-ministic functions on the interval $(0,1]$, such that
$\sup_{r\in(0,1]}|g_{kl}(r)|<\infty$, and each $g_{kl}$ satisfies a Lipschitz
condition piecewise on a finite number of some sub-intervals that partition $(0,1]$.
The matrix $\Sigma(r)=G(r)G(r)'$ is assumed positive definite for all $r$.\\
(ii) The process $(\epsilon_t)$ is $\alpha$-mixing and such that
$E(\epsilon_t\mid \mathcal{F}_{t-1})=0$,
$E(\epsilon_t\epsilon_t'\mid \mathcal{F}_{t-1})=I_d$ and the components $\epsilon_{kt}$ of the process $(\epsilon_t)$ satisfy
$\sup_t\parallel\epsilon_{kt}\parallel_{4\mu}<\infty$ for some
$\mu>1$ and all $k\in\{1,\dots,d\}$.\\

The assumption {\bf A1} generalizes the assumption of Xu and Phillips
(2008) to the multivariate case. From the assumption
$E(\epsilon_t\mid \mathcal{F}_{t-1})=0$, the innovations are
possibly serially dependent. However since $G(r)$ is deterministic
and $E(\epsilon_t\epsilon_t'\mid \mathcal{F}_{t-1})=I_d$, we do not
allow the error process to follow a multivariate GARCH
model. Cavaliere, Rahbek and Taylor (2007)
considered similar volatility structure to
ours. Their
assumption is slightly
different from {\bf A1} in the sense that they do not require a Lipschitz condition and allow for a countable number of jumps.
Boswijk and Zu (2007) allow the matrix
$H_t$ to be possibly stochastic, but requires the volatility process
to be continuous with other additional assumptions, which in particular excludes important cases like abrupt shifts.
Hafner and
Herwartz (2009) assumed no structure on the volatility of the error
process $(u_t)$ and allow for conditional heteroscedasticity. Nevertheless their framework excludes the
use of information on the volatility structure and could result in a loss of
efficiency in the statistical inference of the model. In addition Hafner and
Herwartz (2009) also  assumed
$$\lim_{T\to\infty}T^{-1}\sum_{t=1}^T\Sigma_t=\dot{\Sigma},\quad\mbox{and}\quad\lim_{T\to\infty}T^{-1}\sum_{t=1}^TE\{(\tilde{X}_{t-1}\tilde{X}_{t-1}')\otimes (u_tu_t')\}=W,$$
where $\Sigma_t = E(u_t u_t^\prime)$,
$\tilde{X}_{t-1}=(X_{t-1}',\dots,X_{t-p}')'\in\mathbb{R}^{pd}$ and $W$, $\dot{\Sigma}$ are positive definite matrices, and this could be viewed as too restrictive.
If we suppose that the volatility matrix $H_t$ is constant, we
retrieve the standard homoscedastic case. However the assumption of standard
errors is often considered to be too restrictive for macroeconomic or
financial applications. Indeed many applied studies pointed out
that such data may display unconditional non-stationary volatility (see e.g. Kim and Nelson (1999), Warnock and Warnock (2000) or Batbekh \emph{et al.} (2007)). St\u{a}ric\u{a} and Granger (2005) found that when large samples of stock returns are considered, taking into account shifts for the unconditional volatility instead of assuming a stationary model as a GARCH(1,1) improve the volatility forecasts.

Let us denote by $\theta_0=(\mbox{vec}\:(A_{1})',\dots,\mbox{vec}\:(
A_{p})')'\in\mathbb{R}^{pd^2}$ the vector of the true parameters. The equation (\ref{VAR})
becomes
\begin{eqnarray*}
&& X_t=(\tilde{X}_{t-1}'\otimes I_d)\theta_0+u_t\\&&
u_t=H_t\epsilon_t,\nonumber
\end{eqnarray*}
where we keep the notation $\tilde{X}_{t-1}=(X_{t-1}',\dots,X_{t-p}')'$. Using this expression we first define the OLS estimator
\begin{equation*}
\hat{\theta}_{OLS}=\hat{\Sigma}_{\tilde{X}}^{-1}\mbox{vec}\:\left(\hat{\Sigma}_{X}\right),
\end{equation*}
where
$$\hat{\Sigma}_{\tilde{X}}=T^{-1}\sum_{t=1}^T\tilde{X}_{t-1}\tilde{X}_{t-1}'\otimes
I_d\quad\mbox{and}\quad\hat{\Sigma}_X=T^{-1}
\sum_{t=1}^TX_t\tilde{X}_{t-1}'.$$

Next, let us define the unconditional variance $\Sigma_t:=H_tH_t'$ and the Generalized Least Squares (GLS) estimator that takes into account a time-varying $\Sigma_t$, that is
\begin{equation}\label{GLS}
\hat{\theta}_{GLS}=\hat{\Sigma}_{\tilde{\underline{X}}}
^{-1}\mbox{vec}\:\left(\hat{\Sigma}_{\underline{X}}\right),
\end{equation}
with
$$\hat{\Sigma}_{\tilde{\underline{X}}}=T^{-1}\sum_{t=1}^T\tilde{X}_{t-1}\tilde{X}_{t-1}'
\otimes\Sigma_t^{-1}
\quad\mbox{and}\quad\hat{\Sigma}_{\underline{X}}=T^{-1}
\sum_{t=1}^T\Sigma_t^{-1}X_t\tilde{X}_{t-1}'.$$ 
Note that since $H_t$ is assumed invertible, $\Sigma_t$ is positive definite for all $t$. If we suppose that the volatility matrix $\Sigma_t$ is constant in time, it
is easy to see that $\hat{\theta}_{GLS}=\hat{\theta}_{OLS}$.
However the GLS estimator is in general infeasible since the true volatility
matrix appears in the expression (\ref{GLS}). In the next section we
compare the efficiency of the OLS and GLS estimators.

\section {Asymptotic behaviour of the estimators}
\label{S3}
In order to state the first result of the paper, we need to
introduce the following notations. Since we assumed that $\det
{A}(z)\neq 0$ for all $|z|\leq 1$, it is well known that
\begin{equation}\label{MA}
X_t=\sum_{i=0}^{\infty}\psi_{i}u_{t-i},
\end{equation}
where $\psi_{0}=I_d$ and the components of the $\psi_i$'s are
absolutely summable (see e.g. L\"{u}tkepohl (2005, pp 14-16)). From
the expression (\ref{MA}) we also write
\begin{equation*}\label{MA2}
\tilde{X}_t=\sum_{i=0}^{\infty}\tilde{\psi}_{i}u_{t-i}^p,
\end{equation*}
$u_t^p$ is given by $u_t^p=\mathbf{1}_p\otimes u_t$, where
$\mathbf{1}_p$ is the vector of ones of dimension $p$ and
$$\tilde{\psi}_{i}=\left(
                     \begin{array}{cccc}
                       \psi_i & 0 & 0 & 0 \\
                       0 & \psi_{i-1} & 0 & 0 \\
                       0 & 0 & \ddots & 0 \\
                       0 & 0 & 0 & \psi_{i-p+1} \\
                     \end{array}
                   \right),$$
taking $\psi_j=0$ for $j<0$. Let us define by $\mathbf{1}_{p\times p}$ the $p\times p$ matrix with
components equal to one. The following proposition gives the
asymptotic behavior of the OLS and GLS estimators. For the sake of brevity we only investigate the asymptotic normality, the consistency is in some sense an easier matter and is hence omitted.

\begin{prop}\label{propostu1} If Assumption {\bf A1} holds true, then:
\begin{enumerate}
\item
\begin{equation}\label{res1}
T^{\frac{1}{2}}(\hat{\theta}_{GLS}-\theta_0)\Rightarrow
\mathcal{N}(0,\Lambda_1^{-1}),
\end{equation}
where  $$\Lambda_1=\int_0^1
\sum_{i=0}^{\infty}\left\{\tilde{\psi}_i(\mathbf{1}_{p\times
p}\otimes\Sigma(r))\tilde{\psi}_i'\right\}\otimes\Sigma(r)^{-1}dr$$
is positive definite;
\item
\begin{equation}\label{res2}
T^{\frac{1}{2}}(\hat{\theta}_{OLS}-\theta_0)\Rightarrow
\mathcal{N}(0,\Lambda_3^{-1}\Lambda_2\Lambda_3^{-1}),
\end{equation}
where $$\Lambda_2=\int_0^1
\sum_{i=0}^{\infty}\left\{\tilde{\psi}_i(\mathbf{1}_{p\times
p}\otimes\Sigma(r))\tilde{\psi}_i'\right\}\otimes\Sigma(r)dr$$ and
$$\Lambda_3=\int_0^1
\sum_{i=0}^{\infty}\left\{\tilde{\psi}_i(\mathbf{1}_{p\times
p}\otimes\Sigma(r))\tilde{\psi}_i'\right\}\otimes I_d\:dr$$ are positive definite;

\item The asymptotic variance of $\hat{\theta}_{GLS}$ is smaller than the asymptotic variance of $\hat{\theta}_{OLS}$, that is the matrix
$ \Lambda_3^{-1}\Lambda_2\Lambda_3^{-1} - \Lambda_1^{-1} $
is positive semidefinite.
\end{enumerate}
\end{prop}

If we suppose that the error process is homoscedastic, that is
$\Sigma_t=\Sigma_u$ for all $t$, and since we assumed $E(\epsilon_t\epsilon_t'\mid \mathcal{F}_{t-1})=I_d$, we obtain

$$\Lambda_1=E\left[\tilde{X}_t\tilde{X}_t'\right]\otimes\Sigma_u^{-1},\:\:
\Lambda_2=E\left[\tilde{X}_t\tilde{X}_t'\right]\otimes\Sigma_u
\:\:\mbox{and}\:\:
\Lambda_3=E\left[\tilde{X}_t\tilde{X}_t'\right]\otimes I_d,$$
so that we retrieve the standard result of the iid case (see e.g. L\"{u}tkepohl
(2005, p 74))
\begin{equation}\label{retrieve}
\Lambda_1^{-1}=\Lambda_3^{-1}\Lambda_2\Lambda_3^{-1}
=\{E[\tilde{X}_{t}\tilde{X}_{t}']\}^{-1}\otimes\Sigma_u,
\end{equation}
although here the error process is assumed dependent. Note that in the homoscedastic case the OLS and ALS estimator have the same efficiency.

In
the univariate case ($d=1$), $\Sigma(r)$ belongs to the real line  so that $\Lambda_1$ simplifies to
\begin{equation}\label{phixu} \Lambda_1=
\sum_{i=0}^{\infty}\tilde{\psi}_i\mathbf{1}_{p\times
p}\tilde{\psi}_i, \end{equation} where the $\tilde{\psi}_i$'s are $p\times p$
diagonal matrices. This expression corresponds to the asymptotic covariance
matrix obtained in equation (10) of Xu and Phillips (2008). Moreover,
$$\Lambda_2=\int_0^1\Sigma(r)^2dr
\sum_{i=0}^{\infty}\left\{\tilde{\psi}_i\mathbf{1}_{p\times
p}\tilde{\psi}_i\right\},\quad\Lambda_3=\int_0^1\Sigma(r)dr
\sum_{i=0}^{\infty}\left\{\tilde{\psi}_i\mathbf{1}_{p\times
p}\tilde{\psi}_i\right\},$$ and then we retrieve equation (5) in Xu
and Phillips (2008).

A nice feature of the GLS estimator in the univariate case is
that the covariance matrix of the asymptotic distribution does not
depend on the volatility function $\Sigma(r)$. In the multivariate case the simplification (\ref{phixu})
is still possible if $\Sigma(r)=\sigma^2(r)I_d$, with $\sigma^2(r)$ a scalar function. Nevertheless, we show in Example \ref{ex1} below that
(\ref{phixu}) does not hold in the general multivariate framework and the asymptotic covariance matrix in (\ref{res1}) depends on
the volatility function $\Sigma(r)$. Moreover, our example shows that the covariance matrices in (\ref{res1}) and (\ref{res2}) can be equal in some particular cases of heteroscedasticity but in general they could be very different.

\begin{figure}[h]\!\!\!\!\!\!\!\!\!\!

\vspace*{6.3cm} \hspace*{1.1cm}$\tau_1$ \hspace*{3.2cm}$\sigma_{21}$ \hspace*{5.3cm}$\tau_2$ \hspace*{8.2cm}$\tau_{1}$ \vspace*{0.4 cm}

\protect \includegraphics{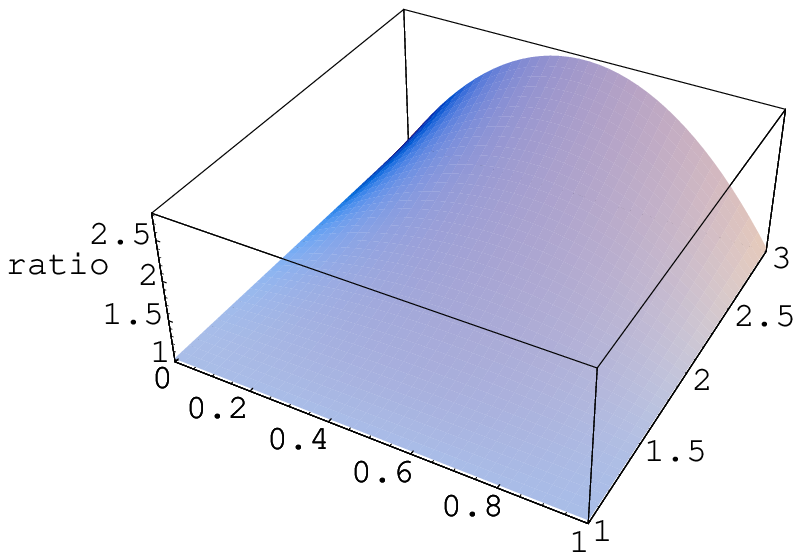} \protect \includegraphics{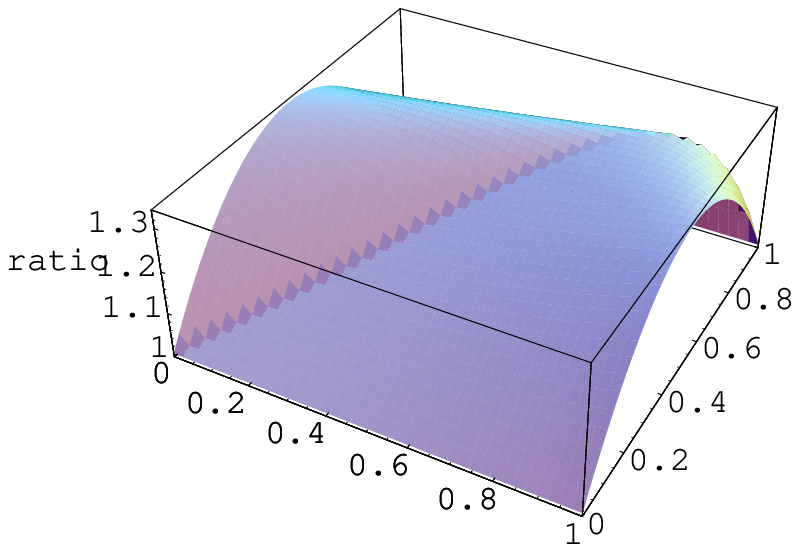} \caption{\label{fig1}
{\footnotesize The ratio
$\mbox{Var}_{as}\left(\hat{\theta}_{2,OLS}\right)/
\mbox{Var}_{as}\left(\hat{\theta}_{2,GLS}\right)$ of Example \ref{ex1}.}}
\end{figure}

\begin{ex}\label{ex1}
{\em Consider the bivariate model (\ref{VAR}) with $p=1$ and

$$A_1=\left(
        \begin{array}{cc}
          a_1 & 0 \\
          0 & a_2 \\
        \end{array}
      \right),
\quad\Sigma(r)=\left(
             \begin{array}{cc}
               \Sigma_1(r) & 0 \\
               0 & \Sigma_2(r) \\
             \end{array}
           \right).$$
In this simple case let us compare the asymptotic variances
$$\mbox{Var}_{as}\left(\hat{\theta}_{2,GLS}\right)=(1-a_1^2)\times
\left(\int_0^1\Sigma_1(r)/\Sigma_2(r)dr\right)^{-1}$$ and
$$\mbox{Var}_{as}\left(\hat{\theta}_{2,OLS}\right)=(1-a_1^2)\times
\left\{\frac{\int_0^1\Sigma_1(r)\Sigma_2(r)dr}{\left(\int_0^1\Sigma_1(r)dr\right)^2}\right\},$$
that is the asymptotic variances of the GLS and OLS estimators of the second component of the vector $\theta_0=(a_1,0,0,a_2)'$ (which corresponds to the element $(2,1)$ of the matrix $A_1$).

First we notice  that $\mbox{Var}_{as}\left(\hat{\theta}_{2,GLS}\right)$ depends on the volatility structure  when $\Sigma_1(r)\neq\Sigma_2(r)$.
In order to illustrate the difference between the variances of
$\hat{\theta}_{2,OLS}$ and $\hat{\theta}_{2,GLS}$, we plot the ratio
\begin{equation}\label{ratiovar}
\mbox{Var}_{as}\left(\hat{\theta}_{2,OLS}\right)/
\mbox{Var}_{as}\left(\hat{\theta}_{2,GLS}\right)
\end{equation}
in Figure \ref{fig1} taking
$$\Sigma_1(r)=\sigma_{10}^2+(\sigma_{11}^2-\sigma_{10}^2)\times \mathbf{1}_{[\tau_1,1]}(r)\quad
\text{and}
 \quad \Sigma_2(r)=\sigma_{20}^2+(\sigma_{21}^2-\sigma_{20}^2)\times \mathbf{1}_{[\tau_2,1]}(r),$$
where  and $\tau_i\in[0,1]$ with $i\in\{1,2\}$. This specification
of the volatility function is inspired by Example 1 of Xu and
Phillips (2008) (see also Cavaliere (2004)). On the left graphic we
take $\tau_1=\tau_2$ and
$\sigma_{10}^2=\sigma_{20}^2=\sigma_{11}^2=1$ but
$\sigma_{21}^2\geq1$, so that only $(X_{2t})$ is heteroscedastic in
general. When $\sigma_{21}^2=1$ or $\tau_1\in\{0, 1\}$, the process
$(X_t)$ is homoscedastic. On the right graphic we take
$\sigma_{10}^2=\sigma_{20}^2=1$ and $\sigma_{11}^2=\sigma_{21}^2=3$
but $\tau_1\neq\tau_2$ in general. When $\tau_1=\tau_2$, we have $\Sigma_1(r)=\Sigma_2(r)$ and hence
we retrieve the case studied in Example 1 of Xu and Phillips (2008).

As expected the ratio (\ref{ratiovar}) is equal to one in the homoscedastic case in
the left graphic. However, departure from this case clearly shows that
the difference between the variances of the two estimators is
increasing with $\sigma_{21}^2$. In the right graphic we can see
that when $\tau_2=0$ or $1$ the ratio in (\ref{ratiovar}) is equal to one
although $(X_t)$ is heteroscedastic. The
variances $\mbox{Var}_{as}(\hat{\theta}_{2,OLS})$ and
$\mbox{Var}_{as}(\hat{\theta}_{2,GLS})$ are different when $\tau_2 \in(0,1)$ and the largest relative difference is attained when we set the volatility shifts in the middle of the sample.}
\end{ex}

It appears that the GLS estimator is more efficient
than the OLS estimator in general when the matrix
$\Sigma_t$ is time-varying. Nevertheless the assumption of known
volatility structure needed to construct the GLS estimator could be unrealistic in practice.
Moreover, the
asymptotic distribution of the GLS estimator depends on the unknown volatility.
In the OLS
estimation approach only the asymptotic distribution of the coefficients estimator depends on the
unknown volatility. In addition,  we can provide simple consistent estimators of $\Lambda_2$ and
$\Lambda_3$, which could be further used for instance to build confidence intervals for the OLS estimators.
For the purpose of estimation of $\Lambda_2$ and
$\Lambda_3$ let us consider the matrices $\Omega_2:=\int_0^1\Sigma(r)^{\otimes2}dr,\quad
\Omega_3:=\int_0^1\Sigma(r)dr$ and denote the OLS residuals by $\hat{u}_t$.

\begin{prop}\label{proposestim} Under Assumption {\bf A1} we have
\begin{equation}\label{om1}
\hat{\Omega}_2:=T^{-1}\sum_{t=2}^T\hat{u}_{t-1}\hat{u}_{t-1}'\otimes\hat{u}_t\hat{u}_t'=\Omega_2+o_p(1),
\end{equation}
\begin{equation}\label{om2}
\hat{\Omega}_3:=T^{-1}\sum_{t=1}^T\hat{u}_t\hat{u}_t'=\Omega_3+o_p(1),
\end{equation}
\begin{equation}\label{L21}
\hat{\Lambda}_2:=T^{-1}\sum_{t=1}^T\tilde{X}_{t-1}\tilde{X}_{t-1}'\otimes\hat{u}_t\hat{u}_t'=\Lambda_2+o_p(1).
\end{equation}
\begin{equation}\label{L3}
\hat{\Lambda}_3:=\hat{\Sigma}_{\tilde{X}}=\Lambda_3+o_p(1),
\end{equation}
\end{prop}

Using (\ref{om1}) and (\ref{om2}) and some additional algebra, we can define alternative consistent estimators
of $\Lambda_2$ and $\Lambda_3$. Indeed, it is shown in the
appendix that


\begin{equation}\label{estim2}
\mbox{vec}\:(\Lambda_2)=\left\{I_{(pd^2)^2}-(\Delta\otimes
I_d)^{\otimes2}\right\}^{-1}\mbox{vec}\:\left(
\begin{array}{cc}
\Omega_2 & 0_{d^2\times(p-1)d^2} \\
0_{(p-1)d^2\times d^2} & 0_{(p-1)d^2\times(p-1)d^2} \\
\end{array}
\right)
\end{equation}
and
\begin{equation}\label{estim3}
\mbox{vec}\:(\Lambda_3)=\left\{I_{(pd^2)^2}-(\Delta\otimes
I_d)^{\otimes2}\right\}^{-1}\mbox{vec}\:\left(
 \begin{array}{cc}
 \Omega_3\otimes I_d & 0_{d^2\times(p-1)d^2} \\
0_{(p-1)d^2\times d^2} & 0_{(p-1)d^2\times(p-1)d^2} \\
\end{array}
\right),
\end{equation}
where $0_{d^2\times(p-1)d^2}$ is the null matrix of dimension $d^2\times(p-1)d^2$ and
$$\Delta=\left(
            \begin{array}{cccc}
              A_1 & \dots & A_{p-1} & A_p \\
              I_d & 0 & \dots & 0 \\
                & \ddots & \ddots & \vdots \\
              0 &  & I_d & 0 \\
            \end{array}
          \right)$$
is a matrix of dimension $pd\times pd$. Therefore replacing $\Omega_2$ and $\Omega_3$ by respectively
$\hat{\Omega}_2$ and $\hat{\Omega}_3$, and the $A_i's$ by their OLS
estimates in the expression of $\Delta$ in (\ref{estim2}) and (\ref{estim3}), we obtain consistent estimators
of $\Lambda_2$ and $\Lambda_3$. These estimators will be denoted by
$\hat{\Lambda}_{2\delta}$ and $\hat{\Lambda}_{3\delta}$, where the subscript $\delta$ refer to the use
of the OLS estimator of $\Delta$. 

\section{Adaptive estimation}
\label{S4}


In the previous section we pointed out that the GLS estimator is generally infeasible in applications. Therefore we consider a feasible weighted estimator obtained using nonparametric estimation of the volatility function. Our approach generalizes the work of Xu and Phillips (2008) to the multivariate case. Let us denote by $A\odot B$ the Hadamard  (entrywise) product of two matrices of same dimension $A$ and $B$. Define the symmetric matrix
$$\check{\Sigma}_t^0=\sum_{i=1}^T w_{ti} \odot \hat{u}_i\hat{u}_i',$$
where, as before the $\hat{u}_i$'s are the OLS residuals and the $kl-$element, $k\leq l$, of the $d\times d$ matrix of weights $w_{ti}$ is given by
$$w_{ti}(b_{kl})= \left(\sum_{i=1}^TK_{ti}(b_{kl})\right)^{-1} K_{ti}(b_{kl}),$$
with $b_{kl}$ the bandwidth and
$$K_{ti} (b_{kl}) =\left\{
              \begin{array}{c}
                K(\frac{t-i}{Tb_{kl}})\quad \mbox{if}\quad t\neq i,\\
                0  \quad\mbox{if}\quad t=i.\\
              \end{array}
            \right.$$
The kernel function $K(z)$ is bounded nonnegative and such that $\int_{-\infty}^\infty K(z)dz=1$. For all $1\leq k\leq l\leq d$ the bandwidth $b_{kl}$ belongs to a range $\mathcal{B}_T = [c_{min} b_T, c_{max} b_T]$ with $c_{min}, c_{max}>0$ some constants and $b_T \downarrow 0$ at a suitable rate that will be specified below.

When using the same bandwidth $b_{kl}\in \mathcal{B}_T$ for all the cells of $\check{\Sigma}_t^0$, since $\hat{u}_i$, $i=1,...,T$ are almost sure linear independent each other, $\check{\Sigma}_t^0$ is almost sure positive definite provided $T$ is sufficiently large. A similar estimator is considered by Boswijk and Zu (2007). When using several bandwidths $b_{kl}$ it is no longer clear that the symmetric matrix $\check{\Sigma}_t^0$ is positive definite. Then we propose to use a regularization of $\check{\Sigma}_t^0$, that is to replace it by the positive definite matrix
\[
\check{\Sigma}_t = \left\{\left(\check{\Sigma}_t^0\right)^2  + \nu_T I_d\right\}^{1/2}
\]
where $\nu_T >0$, $T\geq 1$, is a sequence of real numbers decreasing to zero at a suitable rate that will be specified below. Our simulation experience indicates that in applications with moderate and large samples $\nu_T$ could be even set equal to 0.

In practice the bandwidths $b_{kl}$ can be chosen by minimization of a cross-validation criterion like $$\sum_{t=1}^T\parallel\check{\Sigma}_t-\hat{u}_t\hat{u}_t'\parallel^2,$$
with respect to all $b_{kl}\in\mathcal{B}_T$, $1\leq k\leq l\leq d$, where $\parallel \cdot \parallel $ is some norm for a square matrix, for instance the Frobenius norm that is the square root of the sum of the squares of matrix elements. Our theoretical results below are obtained uniformly with respect to the bandwidths  $b_{kl}\in\mathcal{B}_T$ and this brings a justification for the common cross-validation bandwidth selection approach in the framework we consider. To our best knowledge, this justification is new and hence completes previous procedures of Xu and Phillips (2008) and Boswijk and Zu (2007).

Let us now  introduce the following adaptive least squares (ALS) estimator
\begin{equation*}
\hat{\theta}_{ALS}=\check{\Sigma}_{\tilde{\underline{X}}}
^{-1}\mbox{vec}\:\left(\check{\Sigma}_{\underline{X}}\right),
\end{equation*}
with
$$\check{\Sigma}_{\tilde{\underline{X}}}=T^{-1}\sum_{t=1}^T\tilde{X}_{t-1}\tilde{X}_{t-1}'
\otimes\check{\Sigma}_t^{-1},\quad\mbox{and}\quad\check{\Sigma}_{\underline{X}}=T^{-1}
\sum_{t=1}^T\check{\Sigma}_t^{-1}X_t\tilde{X}_{t-1}'.$$


\vspace{0.3 cm}

\textbf{Assumption A1':} Suppose that all the conditions in Assumption \textbf{A1}(i) hold true.  In addition:

  (i) $\inf_{r\in(0,1]} \lambda_{min}(\Sigma(r)) >0$ where $ \lambda_{min}(\Gamma)$ denotes the smallest eigenvalue of the symmetric matrix $\Gamma$.

 (ii) $\sup_{t} \| \epsilon_{kt} \|_8 <\infty$ for all $k\in\{1,...,d  \}$.

\vspace{0.3 cm}

\textbf{Assumption A2:} \, (i) The kernel $K(\cdot)$ is a bounded density function defined on the real line such that $K(\cdot)$ is nondecreasing on $(-\infty, 0]$ and decreasing on $[0,\infty)$ and $\int_\mathbb{R} v^2K(v)dv < \infty$. The function $K(\cdot)$ is differentiable except a finite number of points and the derivative $K^\prime(\cdot)$  is an integrable function.
Moreover, the Fourier Transform $\mathcal{F}[K](\cdot)$ of $K(\cdot)$ satisfies $\int_{\mathbb{R}}  \left| s \mathcal{F}[K](s) \right|ds <\infty$.

(ii) The bandwidths $b_{kl}$, $1\leq k\leq l\leq d$, are taken in the range $\mathcal{B}_T = [c_{min} b_T, c_{max} b_T]$ with $0< c_{min}< c_{max}< \infty$ and $b_T + 1/Tb_T^{2+\gamma} \rightarrow 0$ as $T\rightarrow \infty$, for some $\gamma >0$.

\vspace{0.3 cm}

Assumption  {\bf A1}' and {\bf A2}(ii) are  natural extensions to the multivariate framework of the assumptions used in Theorem 2 of Xu and Phillips (2008). The conditions on the kernel function are convenient assumptions satisfied by almost all commonly used kernels. These conditions allow us for simpler technical arguments when investigating the rates of convergence uniformly with respect to the bandwidths. The condition on the sequence $b_T$, $T\geq 1$,  is slightly more restrictive than the one imposed by Xu and Phillips (2008) in the univariate case, that is $b_T+1/Tb_T^2\rightarrow 0$, and this is the price we pay for obtaining the results uniformly in the bandwidths in a range $\mathcal{B}_T$.

Let $\Omega_1:=\int_0^1\Sigma(r)\otimes\Sigma(r)^{-1}dr$. In the sequel, we say that a sequence of random matrices $A_T$, $T\geq 1$ is $o_p(1)$ uniformly with respect to (w.r.t.) $b_{kl}\in\mathcal{B}_T$ as $T\rightarrow \infty$ if $\sup_{1\leq k\leq l\leq d} \sup_{b_{kl}\in\mathcal{B}_T} \|\mbox{vec}\:\left(A_T\right)\| \stackrel{P}{\longrightarrow} 0$. The following proposition gives the asymptotic behavior of the
adaptive estimators uniformly w.r.t the bandwidths.

\begin{prop}\label{lemALS}
Under {\bf A1'} and {\bf A2} and provided $T\nu_T^2 \rightarrow 0$, uniformly w.r.t. $b_{kl}\in\mathcal{B}_T$ as $T\rightarrow \infty$
\begin{equation*}
\check{\Lambda}_1:=\check{\Sigma}_{\tilde{\underline{X}}}=\Lambda_1+o_p(1),
\end{equation*}
\begin{equation*}
\check{\Omega}_{1}:=T^{-1}\sum_{t=1}^T\check{\Sigma}_t\otimes\check{\Sigma}_t^{-1}=\Omega_1+o_p(1)
\end{equation*}
and
\begin{equation*}
\sqrt{T}(\hat{\theta}_{ALS}-\hat{\theta}_{GLS})=o_p(1).
\end{equation*}

\end{prop}

\vspace{0.5 cm}

Proposition \ref{lemALS} shows that the ALS and GLS estimators have the
same asymptotic behavior, that is the ALS estimator is consistent in probability and $\sqrt{T}-$asymptotically normal as soon as the GLS estimator has such properties. The results remains true even if the bandwidths $b_{kl}\in\mathcal{B}_T$ are data dependent.

On the other hand, similarly to (\ref{estim2}) and (\ref{estim3}),
\begin{equation}\label{estim1}
\mbox{vec}\:(\Lambda_1)=\left\{I_{(pd^2)^2}-(\Delta\otimes
I_d)^{\otimes2}\right\}^{-1}\mbox{vec}\:\left(
 \begin{array}{cc}
 \Omega_1 & 0_{d^2\times(p-1)d^2} \\
0_{(p-1)d^2\times d^2} & 0_{(p-1)d^2\times(p-1)d^2} \\
\end{array}
\right).
\end{equation}
Then we also obtain an alternative consistent estimator (uniformly w.r.t. $b_{kl}\in\mathcal{B}_T$) $\check{\Lambda}_{1\delta}$
of $\Lambda_1$ by replacing $\Omega_1$ by $\check{\Omega}_1$, and the $A_i's$ by their ALS
estimates in the expression of $\Delta$ in (\ref{estim1}).

\section{Application to the test of the linear Granger causality in mean}
\label{S5}
In this section we propose tests for linear causality in mean in our framework using the
OLS and the adaptive approaches.
Let us consider the subvectors $X_{1t}$ and
$X_{2t}$ such that $X_t=(X_{1t}',X_{2t}')'$ where
$X_{1t}$ is of dimension $d_1<d$, and $d_2=d-d_1$. It is said
that $(X_{2t})$ does not cause linearly  $(X_{1t})$ in mean if
we have
$$EL(X_{1t}\mid X_{1t-1},\dots)=EL(X_{1t}\mid X_{1t-1},X_{2t-1},\dots),$$
where $EL(X_{1t}\mid \dots)$ is the linear conditional
expectation. In our framework since we assumed that $(\epsilon_t)$
is a martingale difference, the linear predictor is optimal.
Therefore we have $EL(X_{1t}\mid \dots)=E(X_{1t}\mid \dots)$,
where $E(X_{1t}\mid \dots)$ is the conditional expectation, and we
simply refer to the linear Granger causality in mean as Granger
causality in mean in the sequel. We test the null hypothesis that
$(X_{2t})$ does not Granger cause $(X_{1t})$ in mean. It is well known that this amounts to test the null hypothesis
that $A_{i,12}=0$ for all $1\leq i \leq p $ versus the alternative that there exists $i\in\{1,\dots,p\}$ such that $A_{i,12}\neq0$,
where the $A_{i,12}$'s are the matrices given by the $d_1$ first
rows and $d_2$ last columns of the $A_i$'s (see
e.g. L\"{u}tkepohl (2005)). Define the block
diagonal matrix $R=diag(C,\dots,C)$ of dimension $pd_1d_2\times
pd^2$, where $C$ is a $d_1d_2\times d^2$-dimensional matrix given by
$$C=\left(
    \begin{array}{cccc}
      0_{d_1\times d_1d}& I_{d_1} & 0_{d_1\times d_2} & 0 \\
      0 & \ddots &  \ddots & 0  \\
      0 & 0 &  I_{d_1} & 0_{d_1\times d_2}  \\
    \end{array}
  \right).$$
The matrix $R$ is such that we have  $R\theta_0=r$ with
$r$ is the null vector of dimension $pd_1d_2$ under the null hypothesis. Therefore the tested
hypotheses can be written as
$$\mathcal{H}_0: R\theta_0=0 \qquad\mbox{vs.}\qquad
\mathcal{H}_1 : R\theta_0\neq0. $$

In this paper we focus on the Wald type tests, because they are the most commonly used tests by the practitioners. 
We first consider the ALS estimator to build tests for Granger causality in mean. Let us introduce the adaptive Wald test statistics
\begin{equation*}
Q_{ALS}=T\hat{\theta}_{ALS}'R'(R\check{\Lambda}_{1}^{-1}R')^{-1}R\hat{\theta}_{ALS}\quad\mbox{and}\quad
Q_{ALS}^{\delta}=T\hat{\theta}_{ALS}'R'(R\check{\Lambda}_{1\delta}^{-1}R')^{-1}R\hat{\theta}_{ALS}.
\end{equation*}
The following proposition gives the asymptotic distribution of the
ALS test statistics as a simple consequence of Proposition \ref{lemALS}. We say that a sequence of random variables $A_T$, $T\geq 1$, converges
in law to a chi-square distribution $\chi^2_{n} $ uniformly w.r.t. $b_{kl}\in\mathcal{B}_T$ as $T\rightarrow \infty$, if there exists a sequence of random variables $\tilde A_T$, $T\geq 1$, independent of $b_{kl}\in\mathcal{B}_T$ such that $\tilde A_T\Rightarrow\chi^2_{n}$ and $A_T - \tilde A_T = o_p(1)$ uniformly w.r.t. $b_{kl}\in\mathcal{B}_T$.

\begin{prop}\label{ALStest} Under the assumptions of Proposition \ref{lemALS}, uniformly w.r.t. $b_{kl}\in\mathcal{B}_T$ as $T\rightarrow \infty$
\begin{equation}\label{ALS1}
Q_{ALS}\Rightarrow\chi^2_{pd_1d_2},
\end{equation}
\begin{equation}\label{ALS2}
Q_{ALS}^{\delta}\Rightarrow\chi^2_{pd_1d_2}
\end{equation}
and
\begin{equation}\label{ALS3}
Q_{ALS}^{\max}=\max\{Q_{ALS},Q_{ALS}^{\delta}\}\Rightarrow\chi^2_{pd_1d_2}.\\\\
\end{equation}
\end{prop}

\vspace{0.4 cm}

Based on Proposition \ref{ALStest} we  propose the following
procedure for testing Granger causality in mean: for a fixed asymptotic level
$\alpha$, reject the null  hypothesis $\mathcal{H}_0$ if
$\chi^2_{pd_1d_2, 1-\alpha} < Q_{ALS}^{\max}$, where $\chi^2_{pd_1d_2, 1-\alpha}$ is the $(1-\alpha)th$ quantile of the $\chi^2_{pd_1d_2}$ law.
Similar procedures could be defined using $Q_{ALS}$ or $Q_{ALS}^{\delta}$ instead  of $Q_{ALS}^{\max}$, but the latter statistic is expected to yield a more powerful test. The tests based on the ALS estimation will be denoted $W_{ALS}$, $W_{ALS}^{\delta}$ and $W_{ALS}^{\max}$ with obvious notations.

Let us now consider the following Wald test statistics based on the OLS estimation
$$Q_{OLS}=T\hat{\theta}_{OLS}'R'(R\hat{\Lambda}_3^{-1}\hat{\Lambda}_2\hat{\Lambda}_3^{-1}R')^{-1}R\hat{\theta}_{OLS},$$
$$Q_{OLS}^\delta=T\hat{\theta}_{OLS}'R'(R\hat{\Lambda}_{3\delta}^{-1}\hat{\Lambda}_{2\delta}\hat{\Lambda}_{3\delta}^{-1}R')^{-1}R\hat{\theta}_{OLS},$$
and the commonly used standard Wald test statistic
$$Q_{S}=T\hat{\theta}_{OLS}'R'(R\hat{J}^{-1}R')^{-1}R\hat{\theta}_{OLS},
\quad\mbox{with}\quad
\hat{J}=\left\{T^{-1}\sum_{t=1}^T\tilde{X}_{t-1}\tilde{X}_{t-1}'\right\}\otimes\hat{\Omega}_3^{-1}.$$
The following proposition gives the asymptotic
behavior of the OLS and standard test statistics.

\begin{prop}\label{propostat} Under {\bf A1} we have as $T\to\infty$
\begin{equation}\label{asympshatols21}
Q_{OLS}\Rightarrow\chi^2_{pd_1d_2},
\end{equation}
\begin{equation}\label{asympshatols0}
Q_{OLS}^{\delta}\Rightarrow\chi^2_{pd_1d_2},
\end{equation}
\begin{equation}\label{asympshatols1}
Q_{OLS}^{\max}=\max\{Q_{OLS},Q_{OLS}^{\delta}\}\Rightarrow\chi^2_{pd_1d_2},
\end{equation}
and
\begin{equation}\label{asymptilols2}
Q_{S}\Rightarrow Z(\delta):=\sum_{i=1}^{pd_1d_2}\kappa_i Z_i^2,
\end{equation}
where the $Z_i$'s are independent ${\cal N}(0,1)$ variables,
$\delta=(\kappa_1,\dots,\kappa_{pd_1d_2})'$ is the vector of the
eigenvalues of the matrix
\begin{equation}\label{cuisine}
\Psi=(RJ^{-1}R')^{-\frac{1}{2}}(R\Lambda_3^{-1}\Lambda_2\Lambda_3^{-1}R')
(RJ^{-1}R')^{-\frac{1}{2}},
\end{equation}
with $$J=\int_0^1
\sum_{i=0}^{\infty}\left\{\tilde{\psi}_i(\mathbf{1}_{p\times
p}\otimes\Sigma(r))\tilde{\psi}_i'\right\}dr\otimes\Omega_3^{-1}.$$
\end{prop}

It is easy to see from (\ref{om2}) and
(\ref{L3}) that $\hat{J}$ is a consistent estimator of $J$.
The results (\ref{asympshatols21}), (\ref{asympshatols0}) and (\ref{asymptilols2}) are direct consequences of
Proposition \ref{propostu1} and \ref{proposestim}, so that the proof
is omitted. In the Appendix we only give the proof of (\ref{asympshatols1}). Similarly to the tests built using the ALS approach,
tests using the results (\ref{asympshatols21}), (\ref{asympshatols0}) and (\ref{asympshatols1}) can be proposed.

When the errors are homoscedastic ($\Sigma_t=\Sigma_{u}$ for all $t$), we obtain
$J=E[\tilde{X}_{t}\tilde{X}_{t}']\otimes\Sigma_{u}^{-1}.$
Recall that in this case we also have $\Lambda_3^{-1}\Lambda_2\Lambda_3^{-1}=\{E[\tilde{X}_{t}\tilde{X}_{t}']\}^{-1}\otimes\Sigma_{u}$, so that we obtain $\Psi=I_{pd_1d_2}$ and hence we retrieve the standard result $Q_{S}\Rightarrow\chi^2_{pd_1d_2}$. However the $\kappa_i$'s in (\ref{asymptilols2}) can be quite different from 1 if the volatility of the errors is not constant as illustrated in the following example.

\begin{figure}[h]\!\!\!\!\!\!\!\!\!\!

\vspace*{6.6cm} \hspace*{4cm}$\tau_1$ \hspace*{3.8cm}$\sigma_{21}$  \vspace*{0.8 cm}

\protect \includegraphics{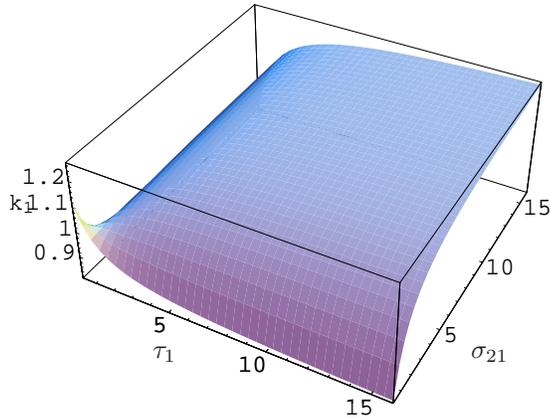}  \caption{\label{exple2}
{\footnotesize The coefficient $\kappa_1$ of Example \ref{ex2}.}}
\end{figure}

\begin{ex}\label{ex2}
{\em Consider the bivariate VAR(1) process $X_t=AX_{t-1}+u_t$ with true parameter $A=0$. Such a model may be used to test Granger causality in mean between the components of an uncorrelated process. Like in Example \ref{ex1}, let us take
$$\Sigma(r)=\left(
              \begin{array}{cc}
                \Sigma_1(r) & 0 \\
                0 & \Sigma_2(r) \\
              \end{array}
            \right).
$$
Suppose that one is interested in testing if $(X_{2t})$ Granger causes $(X_{1t})$ in mean. Then  $R=(0,0,1,0)$ and the matrix $\Psi$ is a scalar such that
$$\Psi=\left(\int_0^1\Sigma_1(r)dr\right)^{-1}\times\left(\int_0^1\Sigma_2(r)dr\right)^{-1}\times\int_0^1\Sigma_1(r)\Sigma_2(r)dr.$$
As a consequence the sum in (\ref{asymptilols2}) reduces to a single term corresponding to the coefficient $\kappa_1=\Psi$. If we suppose that the error process is homoscedastic, we obtain $\kappa_1=1$. However in the general heteroscedastic case we have $\kappa_1\neq1$. To illustrate this let us take
$$\Sigma_1(r)=\sigma_{10}^2+(\sigma_{11}^2-\sigma_{10}^2)r^q$$
and
$$\Sigma_2(r)=\sigma_{20}^2+(\sigma_{21}^2-\sigma_{20}^2)r^q,$$
as in Example 2 of Xu and Phillips (2008). The values of $\kappa_1$ are plotted in Figure \ref{exple2} for $q=1$, $\sigma_{10}=\sigma_{20}=1$ and $\sigma_{11}^2,\sigma_{21}^2\in[0.25, 16]$. It can be seen that in the heteroscedastic case $\kappa_1$ can be quite different from 1 and therefore in this case using the standard Wald procedure based on $Q_S$  for testing if $(X_{2t})$ Granger cause $(X_{1t})$ in mean could be quite a bad idea.}
\end{ex}
The tests based on the results (\ref{asympshatols21}), (\ref{asympshatols0}) and (\ref{asympshatols1}) will be denoted $W_{OLS}$, $W_{OLS}^{\delta}$, $W_{OLS}^{\max}$, and the standard test based on the statistic $Q_S$ and the $\chi^2_{pd_1d_2}$ distribution will be denoted $W_S$.

\section{Monte Carlo experiments}
\label{S6}

The finite sample properties of the OLS, GLS and ALS estimating approaches for VAR analysis are illustrated in this section. Bivariate AR(1) processes are simulated using the model $X_t=AX_{t-1}+u_t$
with

\begin{equation}\label{frenchfuckers}
A=\left(
    \begin{array}{cc}
      a_{11} & a_{12} \\
      a_{21} & a_{22} \\
    \end{array}
  \right)
\quad\mbox{and}\quad u_{t}=H_t\epsilon_t,\quad\quad\epsilon_t\sim\mathcal{N}(0,I_2)\quad\mbox{iid},
\end{equation}
and taking $a_{21}=0.1$ in all the experiments. In addition  if $a_{12}\neq0$ note that $(X_{2t})$ Granger causes
$(X_{1t})$ in mean.
The errors are iid standard Gaussian in the homoscedastic case. We considered this case to study the consequences of the use of the methods for VAR models analysis introduced in this paper while the innovations process is in fact homoscedastic. In the heteroscedastic case the volatility structure is given by
$$\Sigma(r)=\left(
        \begin{array}{cc}
          (1+\gamma_1 r)(1+\rho^2) & \rho(1+\gamma_1 r)^{\frac{1}{2}}(1+\gamma_2 r)^{\frac{1}{2}} \\
          \rho(1+\gamma_1 r)^{\frac{1}{2}}(1+\gamma_2 r)^{\frac{1}{2}} & (1+\gamma_2 r) \\
        \end{array}
      \right),
$$
so that the variances of the error components have a trending behaviour. In the sequel we set $\rho=0.6$ and $\gamma_1=20$, $\gamma_2=\gamma_1/3$.
For the ALS approach the bandwidth is chosen by
cross-validation in a given range as
described in {\bf A2}, and we take $\nu_T=0$ in all the experiments. In the sequel the results for the GLS estimation are given
for comparison only since this method is not feasible in practice. In each experiment $N=1000$ independent trajectories are simulated using (\ref{frenchfuckers}).\\\\ 

%

\begin{figure}[h]\!\!\!\!\!\!\!\!\!\!

\vspace*{1cm} \hspace*{6.0cm}RMSE$\times 10^{2}$

\vspace*{5.6 cm}
\hspace*{11.5cm}$a_{11}$

\protect \includegraphics{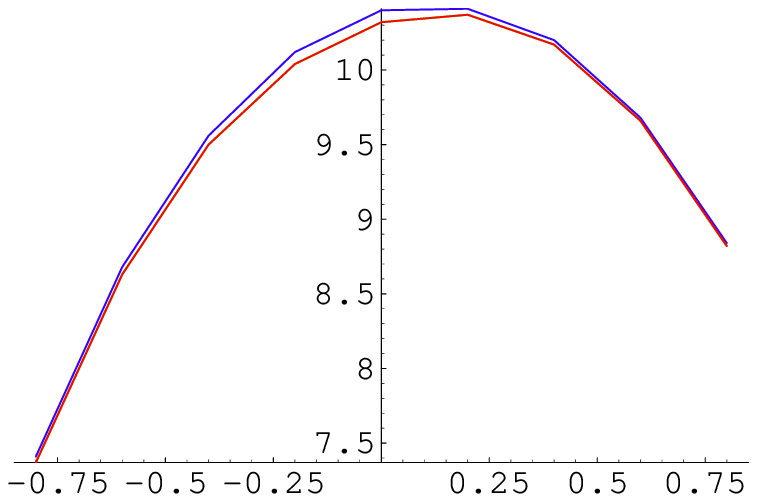}  \caption{\label{fighomo}
{\footnotesize The RMSE of the estimators of the parameters $a_{11}$ over $N=1000$ replications with varying $a_{11}=a_{22}$, $a_{12}=0$ and $a_{21}=0.1$. The errors are homoscedastic and we take $T=100$. The RMSE are displayed in blue for the ALS estimators, in green for the OLS estimators and in red for the GLS estimators.}}
\end{figure}
\begin{figure}[h]\!\!\!\!\!\!\!\!\!\!

\vspace*{1cm} \hspace*{6.0cm}RMSE$\times 10^{2}$

\vspace*{5.6 cm}
\hspace*{11.5cm}$a_{11}$

\protect \includegraphics{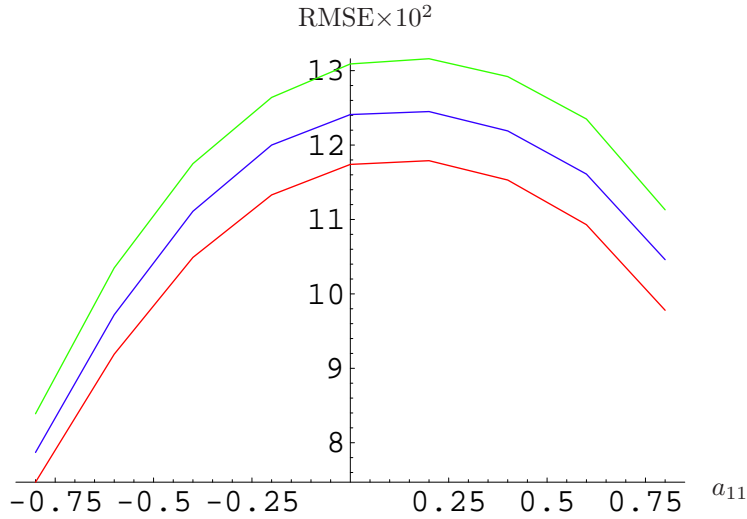} \caption{\label{fig11}
{\footnotesize The RMSE of the estimators of the parameter $a_{11}$ over $N=1000$ replications with varying $a_{11}=a_{22}$, $a_{12}=0$ and $a_{21}=0.1$. We take $\gamma_1=20$, $\rho=0.6$ and $T=100$. The RMSE are displayed in blue for the ALS estimators, in green for the OLS estimators and in red for the GLS estimators.}}
\end{figure}
\begin{figure}[h]\!\!\!\!\!\!\!\!\!\!

\vspace*{1cm} \hspace*{6.0cm}RMSE$\times 10^{2}$

\vspace*{5.9 cm}
\hspace*{11.5cm}$a_{11}$

\protect \includegraphics{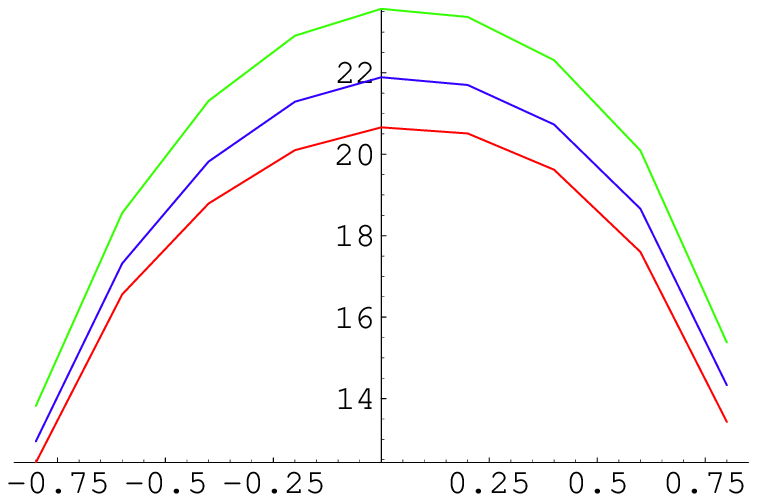} \caption{\label{fig12}
{\footnotesize The same as in Figure \ref{fig11} but for $a_{12}=0$ with varying $a_{22}=a_{11}$ and $a_{21}=0.1$.}}
\end{figure}
\begin{figure}[h]\!\!\!\!\!\!\!\!\!\!

\vspace*{1cm} \hspace*{6.0cm}RMSE$\times 10^{2}$

\vspace*{5.6 cm}
\hspace*{11.5cm}$a_{11}$

\protect \includegraphics{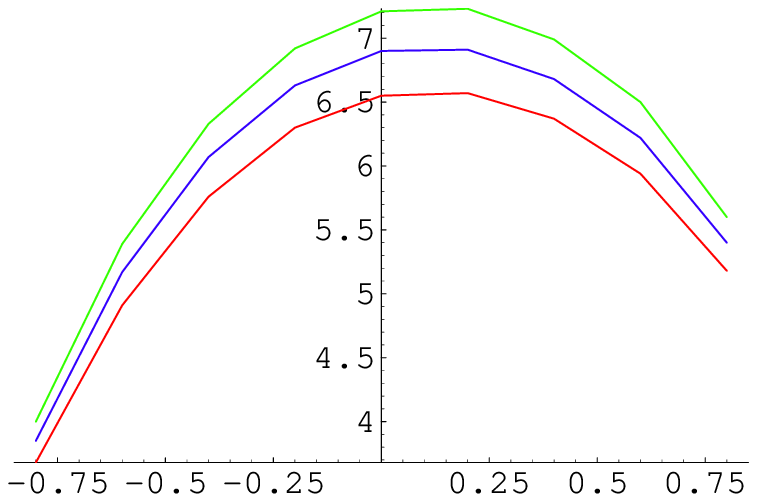}  \caption{\label{fig21}
{\footnotesize The same as in Figure \ref{fig11} but for $a_{21}=0.1$ with varying $a_{22}=a_{11}$ and $a_{12}=0$.}}
\end{figure}
\begin{figure}[h]\!\!\!\!\!\!\!\!\!\!

\vspace*{1.8cm} \hspace*{6.0cm}RMSE$\times 10^{2}$

\vspace*{5.6 cm}
\hspace*{11.5cm}$a_{22}$

\protect \includegraphics{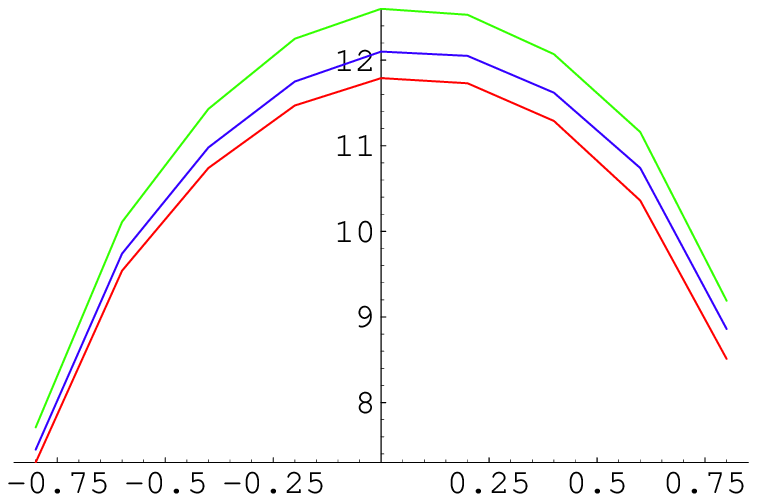} \caption{\label{fig22}
{\footnotesize The same as in Figure \ref{fig11} but for $a_{22}$.}}
\end{figure}

We first examine the properties of the estimation methods presented in the previous sections. The Root Mean Squared Error (RMSE) of the OLS, ALS and GLS methods of the autoregressive parameters is considered in Figures \ref{fighomo} to \ref{fig22}. In these experiments only $a_{11}=a_{22}$ vary and we set $a_{21}=0.1$, $a_{12}=0$. The length of the simulated series is $T=100$. For homoscedastic errors we only give the results for the estimators of $a_{11}$ in Figure \ref{fighomo}.
As expected the infeasible GLS estimation outperforms the other methods in all cases except when the errors are homoscedastic. In this case (Figure \ref{fighomo}) the different estimation methods are equivalent and hence give similar results. This can also be explained by the fact that the ALS based methods choose large bandwidths when the errors are homoscedastic and are then similar to the OLS estimation. However in presence of heteroscedasticity (Figures \ref{fig11} to \ref{fig22}) the ALS procedure clearly better estimate the autoregressive parameters when compared to the OLS estimation.


In this part we study the empirical size of the Wald tests under comparison. Therefore we take $a_{12}=0$, so that $(X_{2t})$ does not causes $(X_{1t})$ in mean. We set $a_{11}=a_{22}=0.2$.
The simulated processes are of lengths $T=50$,
$T=100$, $T=200$ and $T=400$. We test the null hypothesis $a_{12}=0$ at the asymptotic nominal level 5\% in  Tables
\ref{tab1}-\ref{tab2}. Several other values of the autoregressive parameters and specifications of the heteroscedasticity not reported here were experimented, and lead to similar conclusions to that of the presented cases. Since $N=1000$ replications are performed and assuming that the finite sample size of the tests is $5\%$, the relative rejection frequencies should be between the significant
limits 3.65\% and 6.35\% with probability 0.95. Then the relative
rejection frequencies are displayed in bold type when they are
outside these significant limits. 
We first compare the $W_{OLS}$, $W_{S}$, $W_{ALS}$,
and $W_{GLS}$. In Table \ref{tab1} the homoscedastic case is considered. It emerges that the $W_{GLS}$ test outperforms the other tests for $T=50$.
We also remark that the $W_{S}$ test have better results than the $W_{ALS}$ and $W_{OLS}$ tests for $T=50$.
In general the relative rejection frequencies of the different tests are quickly close to the asymptotic nominal level ($T=100$). Therefore in case of doubt of the presence of unconditional heteroscedasticity in the error terms, one can use the ALS and OLS tests without a major loss of efficiency.
In Table \ref{tab2} we use heteroscedastic processes. In accordance with our theoretical results, the $W_S$ test is not valid. The relative rejection frequencies of the $W_{OLS}$, $W_{ALS}$
and $W_{GLS}$ tests converge to the asymptotic nominal level as the samples increase. However we note that the $W_{ALS}$ test have better results
than the $W_{OLS}$ test for $T=50$. It also appears that the infeasible $W_{GLS}$ test have a better control of the error of first kind than the
other tests for $T=50$. From Tables \ref{tab1} and \ref{tab2} the tests $W_{OLS}^{\delta}$, $W_{ALS}^{\delta}$ and $W_{GLS}^{\delta}$
are more liberal than the other tests for small samples.

A further set of Monte Carlo experiments has been conducted to analyze
the empirical power of the studied tests.
We again simulate $N=1000$ independent trajectories of bivariate
AR(1) processes obtained using (\ref{frenchfuckers}), where we set $a_{11}=a_{22}=0.2$.
We take $a_{12}\neq0$, such that the stability condition in (\ref{VAR}) is hold. Hence $(X_{2t})$ is Granger causal in mean for $(X_{1t})$ in this part.
The null
hypothesis $a_{12}=0$ is tested at the asymptotic nominal level 5\%. We only consider samples
of length $T=100$. The results are given in Table \ref{tabpow0} for the homoscedastic case and
in Table \ref{tabpow1} for the heteroscedastic case. 
From our simulation results we remark that the different tests have the same power in the
homoscedastic case. In the heteroscedastic case the $W_{GLS}$ tests are more powerful than
the other tests. It emerges that the ALS tests are more powerful than the OLS tests in
presence of unconditional heteroscedasticity. This can be explained by the fact that the ALS
tests are slightly more sophisticated than the OLS tests. 
We also note a substantial gain of power for the $W_{OLS}^{\max}$, $W_{ALS}^{\max}$
and $W_{GLS}^{\max}$ when compared to the other tests.\\\\ 

We can draw the conclusion that when the process is non stationary but stable, the ALS estimation procedure give significant improvements in the estimation of VAR models when compared to the standard OLS estimation method. We also noted significant improvements of the ALS based tests in the analysis of the linear Granger causality when compared to the OLS based tests in the unconditionally heteroscedastic case. Indeed from our simulation results the ALS tests have a better
control of the error of first kind and a greater ability to detect the linear causality in mean than the OLS based tests in our framework. As expected we found that the standard Wald test is not reliable for the test of autoregressive parameter restrictions when the process is stable but not stationary.


\section{Illustrative example}
\label{S7}

\begin{figure}[h]\!\!\!\!\!\!\!\!\!\!
\vspace*{5.9cm} 


\protect \includegraphics{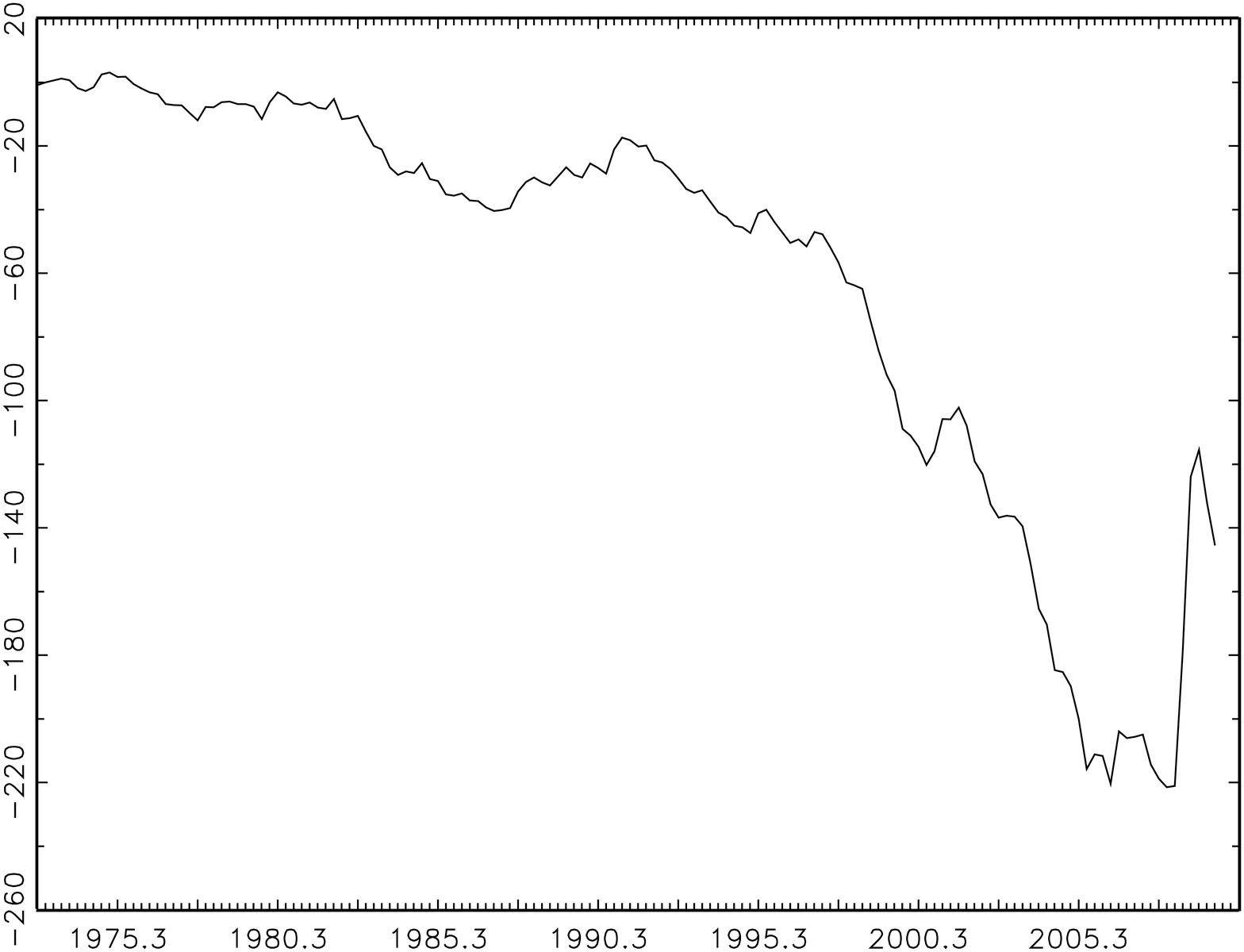} \protect \includegraphics{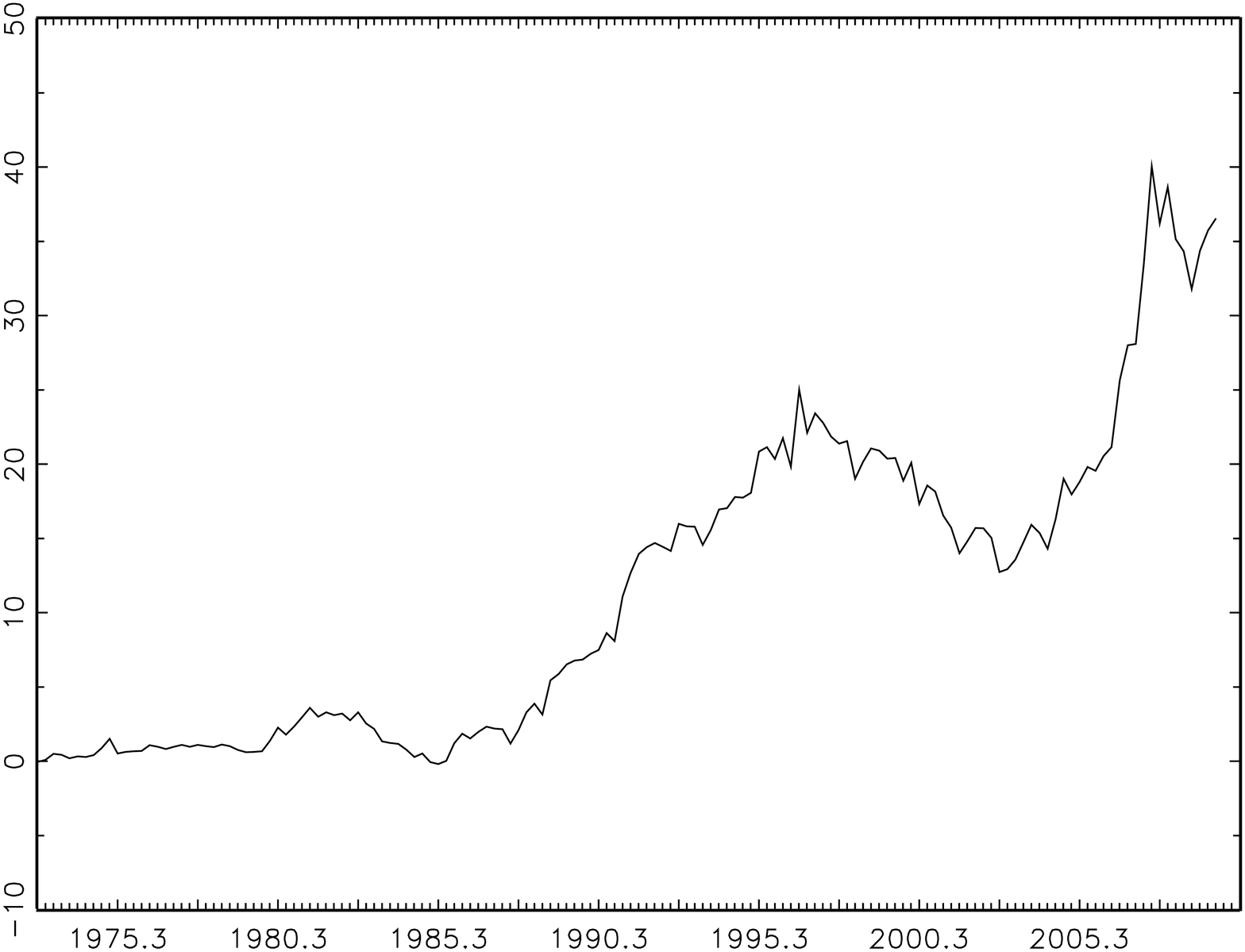} \caption{\label{data}
{\footnotesize The balance on merchandise trade for the U.S. on the left and the balance on services for the U.S. on the right in billions dollars from 1/1/1970 to 10/1/2009, T=160. Data source: The research division of the federal reserve bank of Saint Louis, www.research.stlouis.org.}}
\end{figure}

Now we turn to an example taken from U.S. financial data.
An application to the quarterly U.S. balance on services and balance on merchandise trade in billions of Dollars,
from January 1, 1970 to October 1, 2009 is considered.
The series are available seasonally adjusted from the
website of the research division of the federal reserve bank of Saint
Louis: www.research.stlouisfed.org. The series are plotted in Figure \ref{data}.

\begin{figure}[h]\!\!\!\!\!\!\!\!\!\!
\vspace*{5.9cm} 


\protect \includegraphics{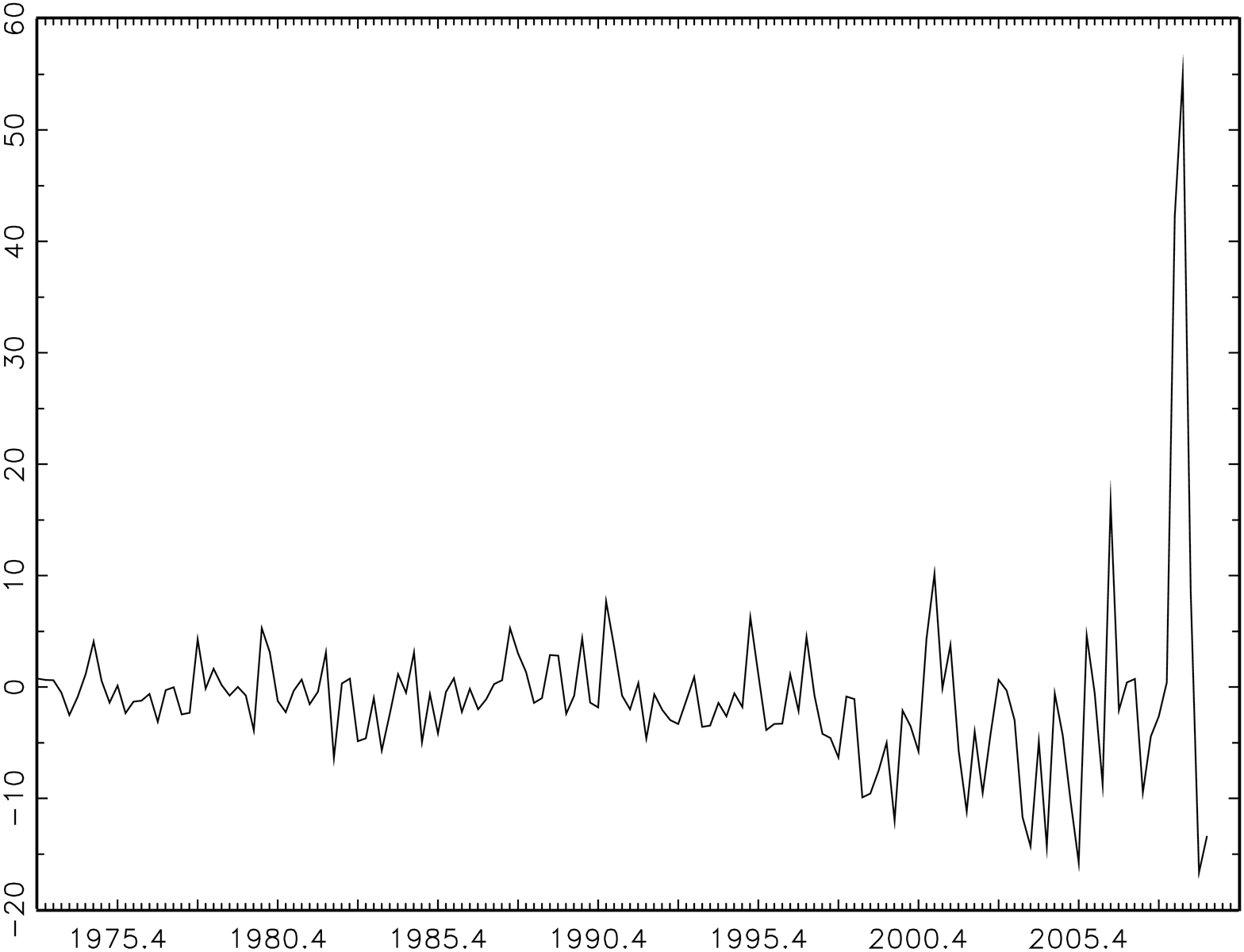} \protect \includegraphics{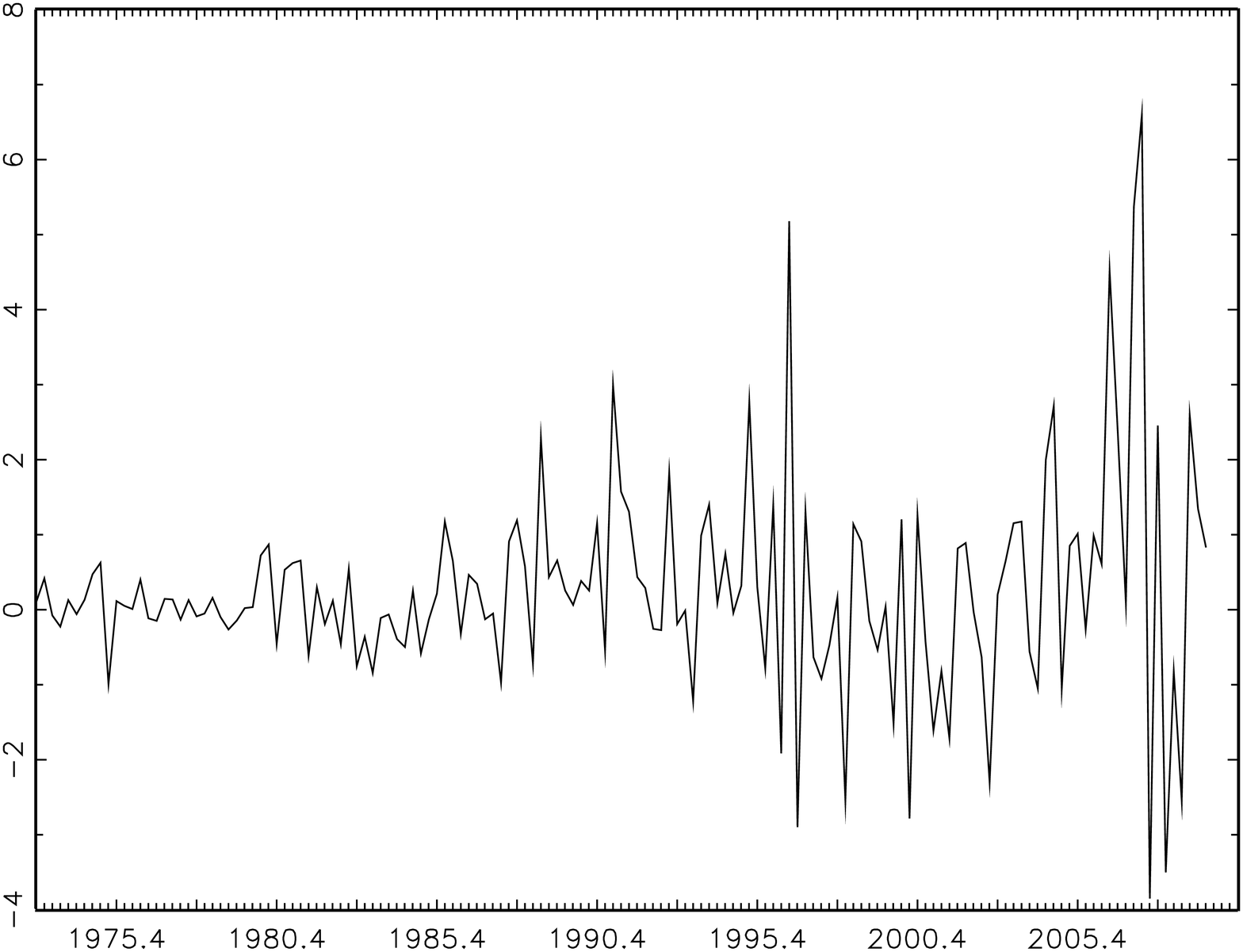} \caption{\label{datadiff}
{\footnotesize The differences of the balance on merchandise trade (on the left) and of the balance on services for the U.S. (on the right).}}
\end{figure}

In a first time we study the properties of the processes. The existence of a unit root for each series is tested using the procedure proposed by Beare (2008). The Augmented Dickey Fuller (ADF) statistic is $3.05$ for the merchandise trade balance data and $7.62$ for the services balance data. These statistics are greater than the 5\% critical value -1.94 of the ADF test. Therefore the the stability hypothesis have to be rejected for both series. In addition we also considered the Kolmogorov-Smirnov (KS) test for homoscedasticity proposed by Cavaliere and Taylor (2008). We found that the KS statistic is $3.05$ for the merchandise trade balance data and is $7.62$ for the services balance data. Since the KS statistics are greater than the 5\% KS critical value 1.36, the homoscedasticity hypothesis is rejected for the studied series. Hence the first differences of the series (plotted in Figure \ref{datadiff}) are considered in the sequel. The length of the series is $T=159$.

\begin{figure}[h]\!\!\!\!\!\!\!\!\!\!
\vspace*{6.9cm} 


\protect \includegraphics{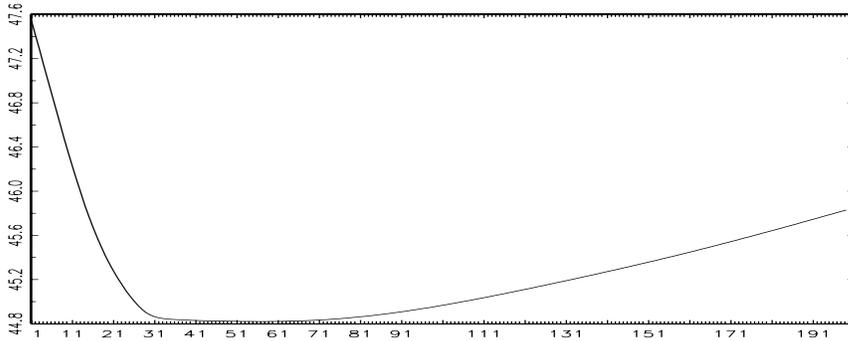}  \caption{\label{crossval}
{\footnotesize The cross validation score for the ALS estimation of the VAR(1) model for the differences of the balance on merchandise trade and on services in the U.S..}}
\end{figure}

\begin{figure}[h]\!\!\!\!\!\!\!\!\!\!
\vspace*{5.9cm} 


\protect \includegraphics{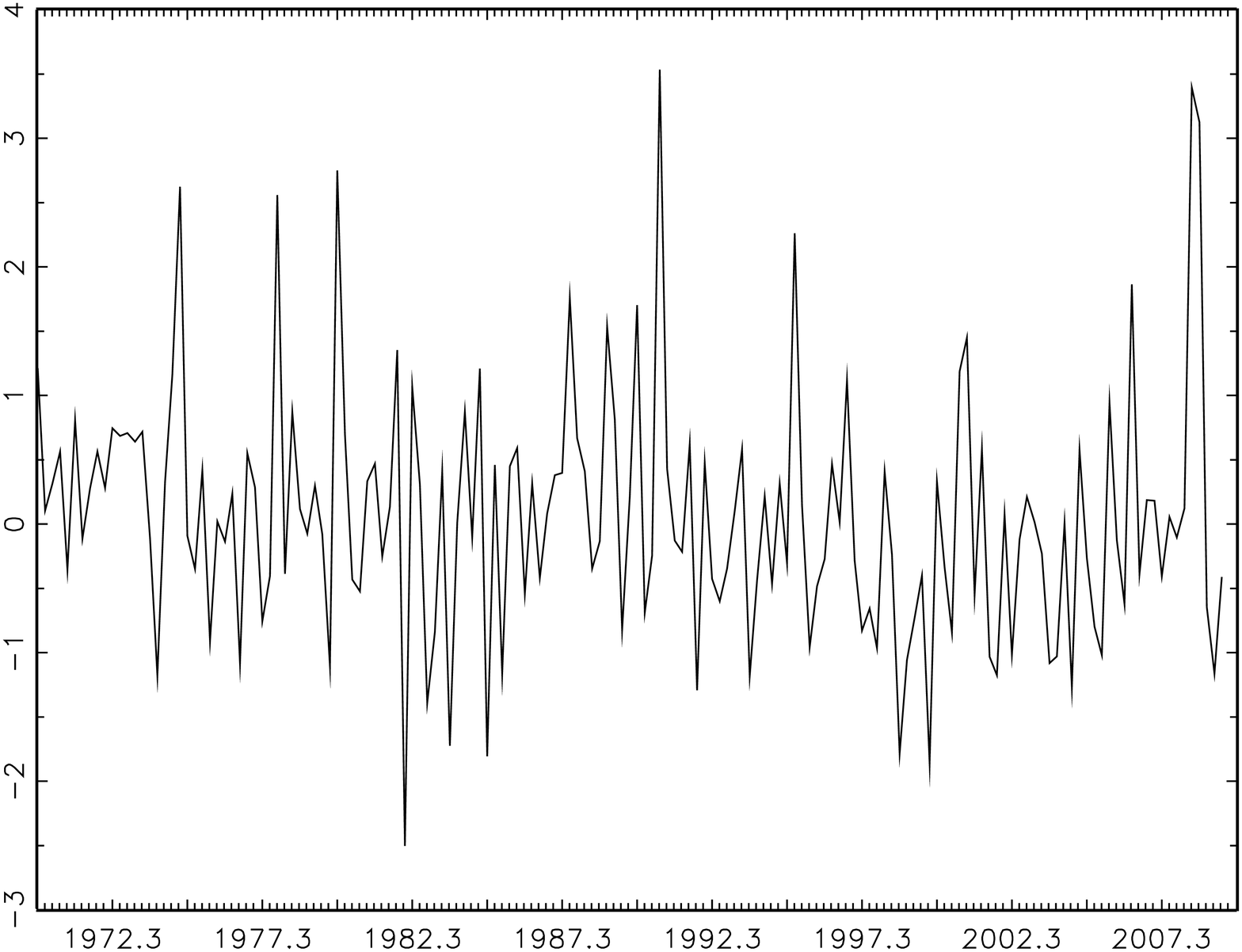} \protect \includegraphics{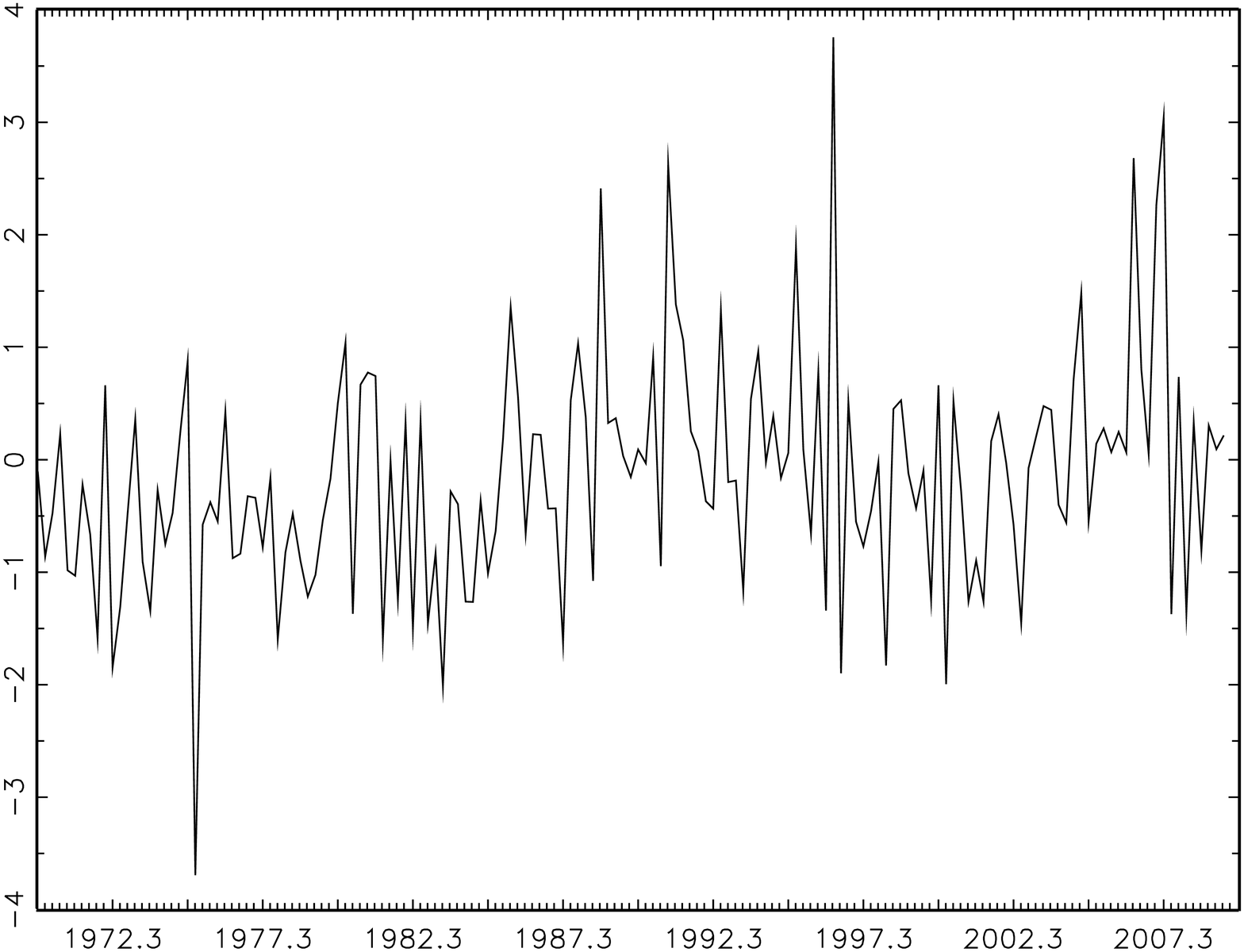} \caption{\label{residals}
{\footnotesize The ALS residuals of a VAR(1) for the differences of the balance on merchandise trade and on services in the U.S.. The first component of the ALS residuals is on the left and the second is on the right.}}
\end{figure}

We fitted a VAR(1) model to the series. The OLS and ALS estimators of the autoregressive parameters are given in Table \ref{estimates}, where the standard deviations obtained using the results (\ref{res1}) and (\ref{res2}) are given into brackets. The standard deviations obtained using the standard result (\ref{retrieve}) are also given to illustrate the case where the practitioner suppose that the processes are homoscedastic. In accordance with our theoretical results we find that the ALS estimators are more precise than the OLS estimators. In view of the results of the KS test one can conclude that the standard deviations based on the homoscedasticity assumption are not reliable.
If a single bandwidth is used for the ALS estimation, $b=7.67\times10^{-2}$ is selected by cross-validation in a given range and using 200 grid points (Figure \ref{crossval}). If several bandwidths are used for the ALS estimation, $b_{11}=1.11\times10^{-2}$, $b_{12}=1.87\times10^{-2}$ and $b_{22}=1.33\times10^{-2}$ are selected in a similar way. From Figure \ref{residals} the ALS residuals seems homoscedastic. Therefore we can deduce that the unconditional heteroscedasticity is well estimated by the adaptive procedure. We also considered the ARCH-LM test of Engle (1982) with different lags in Table \ref{archLM} for testing the presence of ARCH effects in the ALS residuals. It appears that the null hypothesis of conditional homoscedasticity cannot be rejected at the level 5\%. These diagnostics give some evidence that the conditional homoscedasticity assumption on $(\epsilon_t)$ is plausible in our case. However the OLS residuals are clearly heteroscedastic. It is well known that the test of Granger causality in mean strongly depend on the specification of the autoregressive order. Since the standard Box-Pierce procedure is not valid in our framework, we use the modified portmanteau tests proposed by Patilea and Raïssi (2010) to check the goodness-of-fit of the VAR(1) model. We use the Ljung-Box statistic. The tests based on the OLS and ALS estimation are respectively denoted $LB_{m}^{OLS}$ and $LB_{m}^{ALS}$. We also give the result of the standard Ljung-Box test denoted $LB_{m}^{S}$ to illustrate the testing procedure of the Granger causality in mean using only standard tools. The number of autocorrelations used for the portmanteau test statistics is $m=5$ and $15$. From Table \ref{pvalues} it appears that the modified portmanteau tests do not reject the hypothesis of uncorrelated OLS and ALS residuals. The unreliable standard portmanteau tests clearly reject the hypothesis of uncorrelated OLS residuals, so that the practitioner is likely to select a greater autoregressive order. Note that over specified VAR model can entail a loss of efficiency in the test of Granger causality in mean. The reader is referred to Thornton and Batten (1985) for a discussion on the model selection for testing Granger causality in mean.

From the above analysis it appears that the linear dynamics of the series are well described by the VAR(1) model. Hence we can now analyze Granger causality in mean, for instance, from the balance on services to the balance on merchandise trade. In Table \ref{pvaluesgr} we see that the $p$-value of the $W_S$ test is close to zero, so that the null hypothesis of no Granger causality is clearly rejected. However we can remark that the $p$-values of the $W_{ALS}$ and $W_{OLS}$ tests are far from zero. Therefore the null hypothesis is not rejected by the modified tests. These contradictory results can be explained by the fact that the $W_S$ test is not intended to take into account the probable presence of unconditionally heteroscedastic errors in the data, on the contrary of the modified tests which have a larger theoretical basis. It can be also noted in Table \ref{statgr} that the different test statistics are quite different.

\section*{Appendix: Proofs}

We first state some
intermediate results. Define the linear processes
$$\vartheta_t=\sum_{i=0}^{\infty} C_iu_{t-i}^k\quad\mbox{and}\quad
\zeta_t=\sum_{i=0}^{\infty} D_iu_{t-i}^q,$$ where the components of
the $C_i$'s and $D_i$'s are absolutely summable. The vector $u_t^k$
is given by $u_t^k=\mathbf{1}_k\otimes u_t$, where $\mathbf{1}_k$ is the vector of
ones of dimension $k$. Let us introduce $v_t=\mbox{vec}\:(\vartheta_t\zeta_t')$. The following lemmas extend results obtained
in Xu and
Phillips (2008) and Phillips and Xu (2005) to the multivariate case.

\begin{lem}\label{lem0}
\begin{itemize}
\item[(a)]
If $sup_{1\leq t\leq
T}(\parallel\epsilon_{it}\parallel_{2\mu})<\infty$,
$1\leq\mu\leq\infty$, for all $i\in\{1,\dots,d\}$, then we have
$sup_{1\leq t\leq T}(\parallel v_{nt}\parallel_{\mu})<\infty$.
\item[(b)] If $sup_{1\leq t\leq
T}(\parallel\epsilon_{it}\parallel_{4\mu})<\infty$,
$1\leq\mu\leq\infty$, for all $i\in\{1,\dots,d\}$, then we have
$sup_{1\leq t\leq T}(\parallel
\vartheta_{jt}\parallel_{4\mu})<\infty$ for all
$j\in\{1,\dots,kd\}$.
\item[(c)] If $sup_{1\leq t\leq
T}(\parallel\epsilon_{it}\parallel_{4\mu})<\infty$,
$1\leq\mu\leq\infty$, for all $i\in\{1,\dots,d\}$, then we have
$sup_{1\leq t\leq T}(\parallel
\vartheta_{jt-1}\vartheta_{lt-1}u_{j't}u_{l't}\parallel_{\mu})<\infty$
for all $j,j',l,l'\:\in\{1,\dots,kd\}$.
\end{itemize}
\end{lem}

\noindent{\bf Proof of Lemma \ref{lem0}}\quad For the proof of $(a)$
we write
\begin{eqnarray*}
v_t=\sum_{j,l=0}^{\infty}\mbox{vec}\:(C_ju_{t-j}^ku_{t-l}^{'q}D_l')&=&\sum_{j,l=0}^{\infty}(D_l\otimes
C_j)\mbox{vec}\:(u_{t-j}^ku_{t-l}^{'q})
\\&=&\sum_{j,l=0}^{\infty}(D_l\otimes
C_j)(u_{t-l}^q\otimes u_{t-j}^k).
\end{eqnarray*}
Then noting that $E\mid u_{it-l}u_{bt-j}\mid^{\mu}<(E\mid
u_{it-l}\mid^{2\mu}E\mid u_{bt-j}\mid^{2\mu})^{1/2}<\infty$, we have for each component $v_{nt}$ of $v_t\in\mathbb{R}^{kqd^2}$
\begin{eqnarray}\label{vendredi}
E\mid v_{nt}\mid^{\mu}=\parallel v_{nt}\parallel_{\mu}^{\mu}\leq
\left(\sum_{j,l=0}^{\infty}\parallel(D_l\otimes
C_j)\parallel_{\infty} \sup_{i,b\in\{1,\dots,d\}} \parallel u_{it-l}
u_{bt-j}\parallel_{\mu}\right)^{\mu}<\infty,
\end{eqnarray}
where the norm $\parallel .\parallel_{\infty}$ is given by
$\parallel F\parallel_{\infty}=\max_{i}\sum_{j}| f_{ij}|$ for
a square matrix $F$ with obvious notations. Relation
(\ref{vendredi}) is hold since
$\sum_{j,l=0}^{\infty}\parallel(D_l\otimes C_j)\parallel_{\infty}
\leq\sum_{j,l=0}^{\infty}\parallel D_{l}\parallel_{\infty}\parallel
C_j\parallel_{\infty}= \left\{\sum_{l=0}^{\infty}\parallel
D_{l}\parallel_{\infty}\right\} \left\{\sum_{j=0}^{\infty}\parallel
C_{j}\parallel_{\infty}\right\}<\infty$.

For the proof of $(b)$ we write
$$E\mid \vartheta_{jt}\mid^{4\mu}=\parallel\vartheta_{jt}\parallel_{4\mu}^{4\mu}\leq\left(\sum_{l=0}^{\infty}\parallel
D_{l}\parallel_{\infty}\sup_{j\in\{1,\dots,d\}}
\parallel u_{jt-l}\parallel_{4\mu}\right)^{4\mu}<\infty,$$
from the Minkowski inequality, so that $(b)$ hold.\\
For the proof of $(c)$ we have
\begin{eqnarray*}
&&\parallel\vartheta_{jt-1}\vartheta_{lt-1}u_{j't}u_{l't}\parallel_{\mu}^{\mu}=
E\mid
\vartheta_{jt-1}\vartheta_{lt-1}u_{j't}u_{l't}\mid^{\mu}\\&&\leq\left(E\mid
\vartheta_{jt-1}\mid^{4\mu}E\mid\vartheta_{lt-1}\mid^{4\mu}E\mid
u_{j't}\mid^{4\mu}E\mid u_{l't}\mid^{4\mu}\right)^{1/4}<\infty,
\end{eqnarray*}
from the Cauchy-Schwartz inequality, so that $(c)$ hold.
$\quad\square$

\vspace{0.3 cm}

\begin{lem}\label{lem1} Under {\bf A1}
we have
\begin{equation}\label{Gamma}
\lim_{T\to\infty}E\left[\vartheta_{[Tr]-1}\zeta_{[Tr]-1}'\right]=\sum_{i=0}^{\infty}
C_i\left\{\mathbf{1}_{k\times q}\otimes\Sigma(r)\right\}D_i',
\end{equation}
for values $r\in(0,1]$ at which the functions $g_{ij}(r)$ are
continuous, and where $\mathbf{1}_{k\times q}$ is the matrix of ones
of
dimension $k\times q$.
\end{lem}

\vspace{0.3 cm}

\noindent{\bf Proof of Lemma \ref{lem1}}\quad Let us define the
vector $\epsilon_t^k=\mathbf{1}_k\otimes\epsilon_t$. Using the well
known identity $(B\otimes C)(D\otimes F)=(BD)\otimes(CF)$ for
matrices of appropriate dimensions, we have

\begin{eqnarray*}
E[\vartheta_{t-1}\zeta_{t-1}']&=&E\left[\left\{\sum_{i=0}^{\infty}C_iu_{t-i-1}^k\right\}
\left\{\sum_{i=0}^{\infty}D_iu_{t-i-1}^q\right\}'\right]\\&=&
E\left[\sum_{i=0}^{\infty} C_iu_{t-i-1}^ku_{t-i-1}^{'q}D_i'\right]\\&=&
\sum_{i=0}^{\infty} C_iE\left[(\mathbf{1}_k\otimes u_{t-i-1})(\mathbf{1}_q\otimes u_{t-i-1})'\right]D_i'\\&=&
\sum_{i=0}^{\infty} C_iE\left[(\mathbf{1}_k\mathbf{1}_q')\otimes(u_{t-i-1}u_{t-i-1})'\right]D_i'
\\&=& \sum_{i=0}^{\infty}
C_i\left\{\mathbf{1}_{k\times q}\otimes
H_{t-i-1}H_{t-i-1}'\right\}D_i'\\&=& \sum_{i=0}^{\infty}
C_i\left\{\mathbf{1}_{k\times
q}\otimes\Sigma((t-i-1)/T)\right\}D_i'.
\end{eqnarray*}
Now let us write\footnote{Here we make a common abuse of notation because in Assumption \textbf{A1} the matrix-valued function $\Sigma(\cdot)$ is not defined for negative values. To remedy this problem it suffices to extend the function $\Sigma(\cdot)$ to the left of the origin, for instance by setting $\Sigma(r)$ equal to the identity matrix if $r\leq 0$.}
\begin{eqnarray*}
E[\vartheta_{[Tr]-1}\zeta_{[Tr]-1}']&=&\sum_{i=0}^{m}
C_i\left\{\mathbf{1}_{k\times
q}\otimes\Sigma(([Tr]-i-1)/T)\right\}D_i'\\&+&\sum_{i=m+1}^{\infty}
C_i\left\{\mathbf{1}_{k\times
q}\otimes\Sigma(([Tr]-i-1)/T)\right\}D_i',
\end{eqnarray*}
with $m=m(T)\rightarrow\infty$ and $m/T\rightarrow0$. Therefore
noting that $\Sigma(([Tr]-i-1)/T)\rightarrow \Sigma(r)$ as
$T\rightarrow\infty$, with $i<m$, we obtain
$$\sum_{i=0}^{m}
C_i\left\{\mathbf{1}_{k\times
q}\otimes\Sigma(([Tr]-i-1)/T)\right\}D_i'\rightarrow\sum_{i=0}^{\infty}
C_i\left\{\mathbf{1}_{k\times q}\otimes\Sigma(r)\right\}D_i'.$$
Since we assumed that $\sup_{r\in(0,1]}g_{ij}(r)<\infty$ we also
have
$$\sum_{i=m+1}^{\infty}
C_i\left\{\mathbf{1}_{k\times
q}\otimes\Sigma(([Tr]-i-1)/T)\right\}D_i'\rightarrow0$$
as $T\to\infty$, so that we obtain the result (\ref{Gamma}).$\quad\square$

\vspace{0.3 cm}

Recall that we have defined $v_t=\mbox{vec}\:(\vartheta_t\zeta_t')$. We also introduce
$y_t=\mbox{vec}\:(\vartheta_{t-1}\vartheta_{t-1}'\otimes\Sigma_t^{-1}u_tu_t'\Sigma_t^{-1})$
and $z_t=\mbox{vec}\:(\vartheta_{t-1}\vartheta_{t-1}'\otimes
u_tu_t')$.

\begin{lem}\label{precedent} Under {\bf A1} we have
\begin{equation}\label{LLN1}
T^{-1}\sum_{t=1}^Tv_{t}\stackrel{P}{\longrightarrow}\lim_{T\to\infty}T^{-1}\sum_{t=1}^TE(v_{t}).
\end{equation}
\begin{equation}\label{LLN2}
T^{-1}\sum_{t=1}^Ty_{t}\stackrel{P}{\longrightarrow}\lim_{T\to\infty}T^{-1}\sum_{t=1}^TE(y_{t})
=\lim_{T\to\infty}T^{-1}\sum_{t=1}^T
\mbox{vec}\:\left\{E(\vartheta_{t-1}\vartheta_{t-1}')\otimes\Sigma_t^{-1}\right\}.
\end{equation}
\begin{equation}\label{LLN3}
T^{-1}\sum_{t=1}^Tz_{t}\stackrel{P}{\longrightarrow}
\lim_{T\to\infty}T^{-1}\sum_{t=1}^TE(z_t)=\lim_{T\to\infty}T^{-1}\sum_{t=1}^T
\mbox{vec}\:\left\{E(\vartheta_{t-1}\vartheta_{t-1}')\otimes\Sigma_t\right\}.\\
\end{equation}
\end{lem}

\noindent{\bf Proof of Lemma \ref{precedent}}\quad We first show
that $(\vartheta_t)$ is $L^2$-Near Epoch Dependent (NED) on
$(\epsilon_t)$. We write for $l>0$

$$\vartheta_t=\sum_{i=0}^{\infty}C_iu_{t-i}^k=\sum_{i=0}^lC_iu_{t-i}^k+
\sum_{i=l+1}^\infty C_iu_{t-i}^k.$$ Let $\mathcal{F}_{t-l}^{t+l}$ be
the $\sigma$-field generated by
$\{\tilde{u}_{t-l},\dots,\tilde{u}_{t+l}\}$, then
\begin{eqnarray*}
E(\vartheta_{t}\mid\mathcal{F}_{t-l}^{t+l})&=&\sum_{i=0}^lC_iu_{t-i}^k+
\sum_{i=l+1}^\infty C_iE(u_{t-i}^k\mid\mathcal{F}_{t-l}^{t+l})
\end{eqnarray*}
From the Minkowski inequality we have

\begin{eqnarray*}
\parallel\vartheta_{it}-E(\vartheta_{it}\mid\mathcal{F}_{t-l}^{t+l})\parallel_2&\leq&\left\{\sum_{i=1}^\infty\parallel C_{l+i}
\parallel_{\infty}\right\}\sup_{i,j}\parallel u_{jt-l-i}\parallel_2\\&+&\left\{\sum_{i=1}^\infty\parallel C_{l+i}
\parallel_{\infty}\right\}\sup_{i,j}\parallel E(u_{jt-l-i}\mid\mathcal{F}_{t-l}^{t+l})\parallel_2.
\end{eqnarray*}
Since we have

\begin{eqnarray*}
\parallel u_{it-l-1}\parallel_2&=&(E(u_{it-l-1}^2))^{1/2}
\\&=&\left\{E(E(u_{it-l-1}^2\mid\mathcal{F}_{t-l}^{t+l}))\right\}^{1/2}
\\&\geq&\left\{E(E(u_{it-l-1}\mid\mathcal{F}_{t-l}^{t+l})^2)\right\}^{1/2},
\end{eqnarray*}
from the conditional Jensen inequality, and we obtain

\begin{eqnarray*}
\parallel\vartheta_{it}-E(\vartheta_{it}\mid\mathcal{F}_{t-l}^{t+l})\parallel_2&\leq&2\left\{\sum_{i=1}^\infty\parallel C_{l+i}
\parallel_{\infty}\right\}\sup_{i,j}\parallel u_{jt-l-i}\parallel_2.
\end{eqnarray*}
Therefore noting that we have $\sup_{i,j}\parallel
u_{jt-l-i}\parallel_2<\infty$ and since $\sum_{i=1}^\infty\parallel
C_{l+i}
\parallel_{\infty}\rightarrow0$ as $l\rightarrow\infty$, it is clear that $(\vartheta_t)$ is
$L^2$- NED on $(\epsilon_t)$.
Similarly it can be shown that
$(\zeta_t)$ is $L^2$- NED on $(\epsilon_t)$.

From Theorem 17.9 in Davidson (1994), it follows that the process
$\{v_{t}-E(v_{t})\}$ is $L^1$- NED. Therefore since we assumed that
$\epsilon_t$ is $\alpha$-mixing, and using Theorem 17.5 in Davidson
(1994), $\{v_{t}-E(v_{t})\}$ is a $L^1$-mixingale on $(\epsilon_t)$.
In addition using Lemma \ref{lem0} with $\mu=2$, we see that $v_t$
is uniformly integrable. Then from the law of large numbers for
$L^1$-mixingales of Andrews (1988), we obtain (\ref{LLN1}).

For the proof of (\ref{LLN2}) note that
$\mbox{vec}\:\{\vartheta_{t-1}\vartheta_{t-1}'
\otimes\Sigma_t^{-1}u_tu_t'\Sigma_t^{-1}-E(\vartheta_{t-1}\vartheta_{t-1}'
\otimes\Sigma_t^{-1}u_tu_t'\Sigma_t^{-1})\}$ are martingale
differences and uniformly integrable from result (c) of Lemma
\ref{lem0}. Therefore we obtain

$$T^{-1}\sum_{t=1}^T\vartheta_{t-1}\vartheta_{t-1}'
\otimes\Sigma_t^{-1}u_tu_t'\Sigma_t^{-1}\stackrel{P}{\longrightarrow}
\lim_{T\to\infty}T^{-1}\sum_{t=1}^TE(\vartheta_{t-1}\vartheta_{t-1}')
\otimes\Sigma_t^{-1}$$ using similar arguments to that of the proof
of (\ref{LLN1}). The proof of (\ref{LLN3}) is similar to that of
(\ref{LLN2}).$\quad\square$

\vspace{0.3 cm}

\begin{lem}\label{lemtcllln} Under {\bf A1}
we have
\begin{equation}\label{L2}
T^{-1}\sum_{t=1}^T\vartheta_{t-1}\vartheta_{t-1}'\otimes\Sigma_t^{-1}\stackrel{P}{\longrightarrow}
\int_0^1\sum_{i=0}^{\infty}\left\{C_i(\mathbf{1}_{k\times
k}\otimes\Sigma(r))C_i'\right\}\otimes\Sigma(r)^{-1}dr,
\end{equation}

\begin{equation}\label{L1}
T^{-1}\sum_{t=1}^T\vartheta_{t-1}\vartheta_{t-1}'\otimes
I_d\stackrel{P}{\longrightarrow}\int_0^1
\sum_{i=0}^{\infty}\left\{C_i(\mathbf{1}_{k\times
k}\otimes\Sigma(r))C_i'\right\}dr\otimes I_d.
\end{equation}
In addition we also have

\begin{equation}\label{TCL2}
T^{-\frac{1}{2}}\sum_{t=1}^T\mbox{vec}\:(\Sigma_t^{-1}u_t\vartheta_{t-1}')\Rightarrow
\mathcal{N}(0,\Xi_1)
\end{equation}
\begin{equation}\label{TCL1}
T^{-\frac{1}{2}}\sum_{t=1}^T\mbox{vec}\:(u_t\vartheta_{t-1}')\Rightarrow
\mathcal{N}(0,\Xi_2),
\end{equation}

where $$\Xi_1=\int_0^1
\sum_{i=0}^{\infty}\left\{C_i(\mathbf{1}_{k\times
k}\otimes\Sigma(r))C_i'\right\}\otimes\Sigma(r)^{-1}dr,$$ and
$$\Xi_2=\int_0^1
\sum_{i=0}^{\infty}\left\{C_i(\mathbf{1}_{k\times
k}\otimes\Sigma(r))C_i'\right\}\otimes\Sigma(r)dr.$$
\end{lem}

\vspace{0.3 cm}

\noindent{\bf Proof of Lemma \ref{lemtcllln}}\quad For the proof of
(\ref{L2}) we have from Lemma \ref{precedent}

$$T^{-1}\sum_{t=1}^T\vartheta_{t-1}\vartheta_{t-1}'\otimes\Sigma_t^{-1}
\stackrel{P}
{\longrightarrow}\lim_{T\to\infty}T^{-1}\sum_{t=1}^TE(\vartheta_{t-1}\vartheta_{t-1}')\otimes\Sigma_t^{-1}.$$
Now let us denote the discontinuous points of the functions
$g_{ij}(.)$ by $\xi_1,\xi_2,\dots,\xi_q$ where $q$ is a finite
number independent of $T$. We write

\begin{eqnarray*}
&&\lim_{T\to\infty}T^{-1}\sum_{t=1}^TE(\vartheta_{t-1}\vartheta_{t-1}')\otimes\Sigma_t^{-1}
\\&=&\lim_{T\to\infty}\sum_{t=1}^T\int_{t/T}^{(t+1)/T}E(\vartheta_{[Tr]-1}\vartheta_{[Tr]-1}')
\otimes\Sigma_{[Tr]}^{-1}dr+o_p(1)\\&=&\lim_{T\to\infty}\int_{1/T}^{\xi_1}
E(\vartheta_{[Tr]-1}\vartheta_{[Tr]-1}')
\otimes\Sigma_{[Tr]}^{-1}dr+\dots\\&\dots+&\int_{\xi_q}^{(T+1)/T}
E(\vartheta_{[Tr]-1}\vartheta_{[Tr]-1}')
\otimes\Sigma_{[Tr]}^{-1}dr+o_p(1),
\end{eqnarray*}
so that using Lemma \ref{lem1} we obtain

\begin{equation}\label{firstpart}
T^{-1}\sum_{t=1}^T\vartheta_{t-1}\vartheta_{t-1}'\otimes
\Sigma_t^{-1}=\int_0^1
\sum_{i=0}^{\infty}\left\{C_i(\mathbf{1}_{p\times
p}\otimes\Sigma(r))C_i'\right\}\otimes\Sigma(r)^{-1}dr+o_p(1).
\end{equation}
The proof of (\ref{L1}) is similar. For the proof of (\ref{TCL2}),
using the identities $\mbox{vec}\:(ab')=b\otimes a$ and
$\mbox{vec}\:(BJF)=(F'\otimes B)\mbox{vec}\:(J)$ for matrices $B,J
,F$ of appropriate dimensions and vectors $a,b,$ we write

\begin{equation*}
\mbox{vec}\:\left\{ \Sigma_t^{-1}u_t\vartheta_{t-1}'\right\}=
(I_{dp}\otimes\Sigma_t^{-1})(\vartheta_{t-1}\otimes u_t).
\end{equation*}
Then
using again the identity $(B\otimes C)(D\otimes F)=(BD)\otimes(CF)$
we have

\begin{eqnarray*}
\mbox{vec}\:(\Sigma_t^{-1}u_t\vartheta_{t-1}')
\mbox{vec}\:(\Sigma_t^{-1}u_t\vartheta_{t-1}')'&=&
\vartheta_{t-1}\vartheta_{t-1}'
\otimes\Sigma_t^{-1}u_tu_t'\Sigma_t^{-1}.
\end{eqnarray*}
Therefore we obtain the result (\ref{TCL2}) from (\ref{LLN2}), (\ref{L2}) and by
the Lindeberg central limit theorem. Using (\ref{LLN3}), the proof
of (\ref{TCL1}) is
similar to that of (\ref{TCL2}).$\quad\square$

\vspace{0.3 cm}

Now we have the ingredients for proving our results.

\vspace{0.6 cm}

\noindent{\bf Proof of Proposition \ref{propostu1}}\quad 1) and 2). For the
proof of (\ref{res1}) we write using (\ref{VAR}) and (\ref{GLS})
\begin{equation}\label{toimeme}
T^{\frac{1}{2}}(\hat{\theta}_{GLS}-\theta_0)=
\hat{\Sigma}_{\tilde{\underline{X}}}^{-1}\mbox{vec}\:(\hat{\Sigma}_{\underline{Xu}}),
\end{equation}
with
$$\hat{\Sigma}_{\underline{Xu}}=T^{-\frac{1}{2}}\sum_{t=1}^T
\Sigma_t^{-1}u_t\tilde{X}_{t-1}'.$$ Since we have
$\tilde{X}_t=\sum_{i=0}^{\infty}\tilde{\psi}_{i}u_{t-i}^p,$ it
follow from Lemma \ref{lemtcllln} that

\begin{equation*}
\hat{\Sigma}_{\tilde{\underline{X}}}=\int_0^1
\sum_{i=0}^{\infty}\left\{\tilde{\psi}_i(\mathbf{1}_{p\times
p}\otimes\Sigma(r))\tilde{\psi}_i'\right\}\otimes\Sigma(r)^{-1}dr+o_p(1)=\Lambda_1+o_p(1).
\end{equation*}
Using (\ref{TCL2}) we obviously have
$$\hat{\Sigma}_{\underline{Xu}}\Rightarrow \mathcal{N}(0,\Lambda_1),$$
so that we obtain the result (\ref{res1}).

For the proof of (\ref{res2}) we write similarly to (\ref{toimeme})
$$T^{\frac{1}{2}}(\hat{\theta}_{OLS}-\theta_0)=
\hat{\Sigma}_{\tilde{X}}^{-1}\mbox{vec}\:(\hat{\Sigma}_{Xu}),$$ with
$$\hat{\Sigma}_{Xu}=T^{-\frac{1}{2}}\sum_{t=1}^Tu_t\tilde{X}_{t-1}'.$$
From (\ref{L1}) and (\ref{TCL1}) we write
\begin{equation}\label{Hilbert}
\hat{\Sigma}_{\tilde{X}}=\int_0^1
\sum_{i=0}^{\infty}\left\{\tilde{\psi}_i(\mathbf{1}_{p\times
p}\otimes\Sigma(r))\tilde{\psi}_i'\right\}dr\otimes
I_d+o_p(1)=\Lambda_3+o_p(1),
\end{equation}
and
$$\mbox{vec}\:(\hat{\Sigma}_{Xu})\Rightarrow\mathcal{N}(0,\Lambda_2),$$
with
$$\Lambda_2=\int_0^1
\sum_{i=0}^{\infty}\left\{\tilde{\psi}_i(\mathbf{1}_{p\times
p}\otimes\Sigma(r))\tilde{\psi}_i'\right\}\otimes\Sigma(r)dr,$$ so
that we obtain the result (\ref{res2}).

In this part we show that $\Lambda_3$ is positive definite. To this
aim it suffices to show that the matrix
$\tilde{\Lambda}_3=\sum_{i=0}^{\infty}\tilde{\psi}_i(\mathbf{1}_{p\times
p}\otimes\Sigma(r))\tilde{\psi}_i'$ is positive definite for all $r$.
Let us consider a $pd$-dimensional vector $\lambda\neq0$. If
$\tilde{\Lambda}_3$ is not positive definite we have
$$\sum_{i=0}^{\infty}\lambda'\tilde{\psi}_i(\mathbf{1}_{p\times
p}\otimes\Sigma(r))\tilde{\psi}_i'\lambda=\sum_{i=0}^{\infty}\tilde{\lambda}_i'\tilde{\lambda}_i=\sum_{i=0}^{\infty}\tilde{\lambda}_i^2=0,$$
where
$\tilde{\lambda}_i'=(\lambda_1'\psi_iG(r),\dots,\lambda_p'\psi_{i-p+1}G(r))$
with obvious notations. Therefore we have $\tilde{\lambda}_i=0$ for
all $i\in \mathbb{N}$. First consider $\tilde{\lambda}_0$. In this
case we have $\psi_0=I_d$ and
$\psi_{-1}=\psi_{-2}=...=\psi_{i-p+1}=0$. Since we assumed that $\Sigma(r)$ is positive definite
we can deduce that
$\lambda_1=0$. Similarly $\tilde{\lambda}_1$ implies that
$\lambda_2=0$, $\tilde{\lambda}_2$ implies that $\lambda_3=0$ and so
on. Thus $\lambda=0$, which shows that $\Lambda_3$ is positive definite. Using similar arguments and since the Kronecker product of two positive definite matrices is positive definite, it can be shown that the matrices
$\Lambda_1$ and $\Lambda_2$ are positive definite.

3) Using the Cholesky decomposition for positive semidefinite matrix we can write
$$\sum_{i=0}^{\infty}\left\{\tilde{\psi}_i(\mathbf{1}_{p\times
p}\otimes\Sigma(r))\tilde{\psi}_i'\right\} = Z(r)Z' (r),$$
and let
\[
B_k(r) =\{ Z(r) \otimes \Sigma^{k-3/2} (r) \}', \quad k=1,2\quad r\in(0,1].
\]
Then, by the properties of the Kronecker product we have
$$\Lambda_k=\int_0^1 B'_k(r) B_k(r) dr,\quad k=1,2,$$
and
$$\Lambda_3=\int_0^1 B'_2(r) B_1(r) dr = \int_0^1 B'_1(r) B_2(r) dr.$$
Define
\[
\Lambda = \left\{ \int_0^1 B'_2(r) B_2(r) dr \right\}^{-1} \int_0^1 B'_2(r) B_1(r) dr = \Lambda_2^{-1} \Lambda_3.
\]
Following the idea of Lavergne (2008) we can write
\begin{eqnarray*}
0 &\ll & \int_0^1 \{ B_1(r) - B_2(r) \Lambda \}'  \{ B_1(r) - B_2(r) \Lambda  \} dr\\
&=& \Lambda_1 - \Lambda' \int_0^1 B'_2(r) B_1(r) dr  - \int_0^1 B'_1(r) B_2(r) dr \Lambda
 +  \Lambda' \Lambda_2 \Lambda\\
&=& \Lambda_1 - \Lambda_3 \Lambda_2^{-1} \Lambda_3
\end{eqnarray*}
and this prove the stated result. Notice that the equality between the two asymptotic variance holds if and only if $B_1(r) = B_2(r) \Lambda$ for almost all $r\in(0,1]$.
$\quad\square$

\vspace{0.3 cm}

\noindent{\bf Proof of Proposition \ref{proposestim}} For the proof of
(\ref{om2}) we write

\begin{eqnarray}\label{time}
T^{-1}\sum_{t=1}^T\hat{u}_t\hat{u}_t'&=&T^{-1}\sum_{t=1}^Tu_tu_t'-
\left[\sum_{i=1}^p\left\{T^{-1}\sum_{t=1}^Tu_tX_{t-i}'\right\}(\hat{A}_i^{OLS}-A_i)'\right]
\nonumber\\&-&\left[\sum_{i=1}^p(\hat{A}_i^{OLS}-A_i)\left\{T^{-1}\sum_{t=1}^TX_{t-i}u_t'\right\}\right]
\nonumber\\&-&\left[\sum_{i=1}^p(\hat{A}_i^{OLS}-A_i)\left\{T^{-1}\sum_{t=1}^TX_{t-i}X_{t-i}'\right\}
\nonumber(\hat{A}_i^{OLS}-A_i)'\right],
\\&=&c_1+c_2+c_3+c_4,
\end{eqnarray}
with obvious notations. Using similar arguments to that of the proof
of (\ref{TCL1}), it is easy to see that we have

$$T^{-1}\sum_{t=1}^Tu_tX_{t-i}'=O_p(T^{-\frac{1}{2}})\quad\mbox{and}\quad T^{-1}\sum_{t=1}^TX_{t-i}'u_t=O_p(T^{-\frac{1}{2}}).$$
Then since $\hat{A}_i^{OLS}-A_i=O_p(T^{-\frac{1}{2}})$, we write
$c_2=o_p(1)\quad\mbox{and}\quad c_3=o_p(1)$. From relation
(\ref{L1}), it is also easy to see that we have $c_4=o_p(1)$. Let us
define $w_t=\mbox{vec}(u_tu_t'-vec(G(t/T)G(t/T)')$. Since
$\{w_t,\mathcal{F}_{t-1}\}$ is $\alpha$-mixing by Theorem 14.1 in
Davidson (1994), and $E\parallel w_t\parallel ^2<\infty$ by {\bf
A1}, we have by the law of large numbers for $L^1$-mixingales

\begin{eqnarray*}
T^{-1}\sum_{t=1}^T\mbox{vec}(u_tu_t')&=&\lim_{T\to\infty}T^{-1}\sum_{t=1}^TE\{\mbox{vec}(u_tu_t')\}+o_p(1)
\\&=&\lim_{T\to\infty}T^{-1}\sum_{t=1}^T\mbox{vec}(G(t/T)G(t/T)')+o_p(1)\\&=&\mbox{vec}\int_0^1\Sigma(r)dr+o_p(1),
\end{eqnarray*}
and we obtain (\ref{om2}). For the proof of (\ref{om1}) we have similarly to (\ref{time})

$$T^{-1}\sum_{t=2}^T\hat{u}_{t-1}\hat{u}_{t-1}'\otimes\hat{u}_t\hat{u}_t'
=T^{-1}\sum_{t=2}^Tu_{t-1}u_{t-1}'\otimes u_tu_t'+o_p(1).$$
From the Cauchy-Schwartz inequality and by Assumption {\bf A1} we have

$$E\mid u_{it-1}u_{jt-1}u_{kt}u_{lt}\mid^{\mu}<
\{E(u_{it-1})^{4\mu}E(u_{jt-1})^{4\mu}E(u_{kt})^{4\mu}E(u_{lt})^{4\mu}\}^{\frac{1}{4}}<\infty.$$
Then using again the law of large numbers for $L^1$-mixingales and since

$$E(u_{t-1}u_{t-1}'\otimes u_tu_t')=E(u_{t-1}u_{t-1}'\otimes E(u_tu_t'\mid\mathcal{F}_{t-1}))=\Sigma_{t-1}\otimes\Sigma_t,$$
we write

\begin{eqnarray*}T^{-1}\sum_{t=2}^Tu_{t-1}u_{t-1}'\otimes u_tu_t'&=&\lim_{T\to\infty}T^{-1}\sum_{t=2}^T\Sigma_{t-1}\otimes\Sigma_t+o_p(1)
\\&=&\lim_{T\to\infty}T^{-1}\sum_{t=2}^T\Sigma(t-1/T)\otimes\Sigma(t/T)+o_p(1).
\end{eqnarray*}
Finally noting that
$$\lim_{T\to\infty}T^{-1}\sum_{t=2}^T\Sigma(t-1/T)\otimes\Sigma(t/T)=\int_0^1\Sigma(r)^{\otimes2}dr+o_p(1).$$
we obtain (\ref{om1}). The proof of (\ref{L3}) follows from (\ref{Hilbert}). For the proof of (\ref{L21}), we write as above

$$T^{-1}\sum_{t=1}^T\tilde{X}_{t-1}\tilde{X}_{t-1}'\otimes\hat{u}_t\hat{u}_t'=
T^{-1}\sum_{t=1}^T\tilde{X}_{t-1}\tilde{X}_{t-1}'\otimes u_tu_t'+o_p(1),$$
so that we obtain the desired result from similar arguments used for the proof of (\ref{TCL2}). $\quad\square$

\vspace{0.3 cm}

\noindent{\bf Proof of (\ref{estim2}), (\ref{estim3}) and (\ref{estim1})}\quad We
only prove (\ref{estim3}). The proofs of (\ref{estim2}) and (\ref{estim1}) are similar.
From the proof of Theorem \ref{propostu1} we have
\begin{equation}\label{memelimit}
\hat{\Sigma}_{\tilde{X}}=\Lambda_3+o_p(1).
\end{equation}
Using Lemma \ref{precedent} we also obtain
\begin{equation*}
\mbox{vec}\:\{\hat{\Sigma}_{\tilde{X}}\}\stackrel{P}
{\longrightarrow}\lim_{T\to\infty}T^{-1}\sum_{t=1}^T\mbox{vec}\:
\{E(\tilde{X}_{t-1}\tilde{X}_{t-1}')\otimes
I_d\}.
\end{equation*}
Straightforward computations show that
$$\mbox{vec}\:\{E(\tilde{X}_{t-1}\tilde{X}_{t-1}')\otimes I_d\}=
\sum_{i=0}^{\infty}\{(\Delta\otimes
I_d)^{\otimes2}\}^i\mbox{vec}\:\left[\left(
                                                                    \begin{array}{cc}
                                                                      \Sigma(\frac{t-i-1}{T}) & 0 \\
                                                                      0 & 0 \\
                                                                    \end{array}
                                                                  \right)
\otimes I_d\right].$$ Then considering similar arguments used in the
proof of Lemma \ref{lem1} we write
\begin{eqnarray*}
&&\lim_{T\to\infty}\mbox{vec}\:\{E(\tilde{X}_{[Tr]-1}\tilde{X}_{[Tr]-1}')\otimes
I_d\}\\&=& \left\{I_{(pd^2)^2}-(\Delta\otimes
I_d)^{\otimes2}\right\}^{-1}\mbox{vec}\:\left[\left(
                                                                    \begin{array}{cc}
                                                                      \Sigma(r)\otimes I_d & 0 \\
                                                                      0 & 0 \\
                                                                    \end{array}
                                                                  \right)\right],
\end{eqnarray*}
so that we obtain
\begin{equation*}
\mbox{vec}\:\{\hat{\Sigma}_{\tilde{X}}\}=\left\{I_{(pd^2)^2}-(\Delta\otimes
I_d)^{\otimes2}\right\}^{-1}\mbox{vec}\:\left[\left(
                                                                    \begin{array}{cc}
                                                                      \int_0^1\Sigma(r)dr\otimes I_d & 0 \\
                                                                      0 & 0 \\
                                                                    \end{array}
                                                                  \right)\right]+o_p(1).
\end{equation*}
by using similar arguments of the proof of (\ref{L2}). Therefore the
result (\ref{estim3}) follow from (\ref{memelimit}). $\quad\square$

\vspace{0.3 cm}

\noindent{\bf Proof of Proposition \ref{lemALS}} In the following, $c$, $C$, ... denote constants with possibly different values from line to line. First, let us focus on the asymptotic equivalence between $\hat{\theta}_{ALS}$ and $\hat{\theta}_{GLS}$ uniformly w.r.t. the bandwidths $b_{kl}\in\mathcal{B}_T$. We extend the arguments of Theorem 2 in Xu and Phillips (2008). Consider the notation
\[
A(\Gamma)=T^{-1}\sum_{t=1}^T\tilde{X}_{t-1}\tilde{X}_{t-1}'
\otimes\Gamma_t^{-1},\quad\mbox{and}\quad a(\Gamma)=T^{-1/2}
\sum_{t=1}^T\Gamma_t^{-1}u_t\tilde{X}_{t-1}'.
\]
Then
\begin{eqnarray*}
\sqrt{T}(\hat{\theta}_{ALS}-\hat{\theta}_{GLS}) &=& A(\check{\Sigma})^{-1} \mbox{vec}\:\left(a(\check{\Sigma})\right) - A(\Sigma)^{-1} \mbox{vec}\:\left(a(\Sigma)\right)\\
&=& A(\check{\Sigma})^{-1} \left\{ \mbox{vec}\:\left(a(\check{\Sigma})\right)
- \mbox{vec}\:\left(a(\Sigma)\right) \right\}\\ &&  - A(\Sigma)^{-1}\left\{ A(\check{\Sigma}) - A(\Sigma) \right\} A(\check{\Sigma})^{-1} \mbox{vec}\:\left(a(\Sigma)\right).
\end{eqnarray*}
By our result (\ref{L2}), $ A(\Sigma) \stackrel{P}{\longrightarrow} \Lambda_1$ which is positive definite. Moreover, $a(\Sigma)$ is bounded in probability by Markov's inequality, Lemma \ref{lem0}-a) considered with $\mu \geq 2$ and the linear processes $\vartheta_t = u_t $ and $\zeta_t = \tilde{X}_{t-1}$, and the fact that $\Sigma_t^{-1}$ is bounded. Hence, like in the proof of Theorem 2 of Xu and Phillips (2008), to prove that $\sqrt{T}(\hat{\theta}_{ALS}-\hat{\theta}_{GLS}) = o_p(1)$, uniformly w.r.t. $b_{kl}\in\mathcal{B}_T$, it suffices to check
\begin{equation}\label{qsqs}
A(\check{\Sigma}) - A(\Sigma) = o_p(1) \quad \text{and} \quad a(\check{\Sigma}) - a(\Sigma) = o_p(1),
\end{equation}
uniformly w.r.t. $b_{kl}\in\mathcal{B}_T$. As a direct by-product we also obtain $\check{\Lambda}_1 - \Lambda_1=o_p(1)$ uniformly w.r.t. the bandwidths $b_{kl}$. Let us define
\begin{equation}\label{sigma_rond}
\stackrel{\circ}{\Sigma}_t = \sum_{i=1}^T w_{ti}\odot {u}_i{u}_i'\quad \text{and} \quad \bar \Sigma_t = \sum_{i=1}^T w_{ti} \odot \Sigma_i,
\end{equation}
and, following Xu and Phillips (see also Robinson, 1987), notice that the results in (\ref{qsqs}) are  consequences of the following eight rates obtained uniformly w.r.t. $b_{kl}\in\mathcal{B}_T$: (a) $a(\check{\Sigma}^0) - a(\stackrel{\circ}{\Sigma}) = o_p(1)$; (a') $  a(\check{\Sigma}) - a(\check{\Sigma}^0)= o_p(1)$; (b) $a(\stackrel{\circ}{\Sigma}) - a(\bar \Sigma) = o_p(1)$; (c) $a(\bar \Sigma) - a(\Sigma) = o_p(1)$; (d) $A(\check{\Sigma}^0) - A(\stackrel{\circ}{\Sigma}) = o_p(1)$; (d') $A(\check{\Sigma}) -   A(\check{\Sigma}^0)= o_p(1)$; (e) $A\left(\stackrel{\circ}{\Sigma}\right) - A(\bar \Sigma) = o_p(1)$; (f) $A(\bar \Sigma) - A(\Sigma) = o_p(1)$.
In this proof  the norm $\| \cdot\|$ is the Frobenius norm which in particular is a sub-multiplicative norm, that is $\| AB\| \leq \| A\| \| B\|$, and for a positive definite matrix $A$, $\|A\| \leq C[\lambda_{min} (A) ]^{-1}$ with $C$ a constant depending only on the dimension of $A$. Moreover, $\|A \otimes B \| = \| A\| \|B \|$. To simplify notation, let $b$ denote the $d(d+1)$ vector of bandwidths $b_{kl}$, $1\leq k\leq l\leq d$. Below we will simply write \emph{uniformly w.r.t. $b$} instead of \emph{uniformly w.r.t. $b_{kl}$, $1\leq k\leq l\leq d$}, and  $\sup_b$ instead of $\sup_{b_{kl}\in\mathcal{B}_T, 1\leq k\leq l\leq d}$.

(a) Using the identity $A^{-1} - B^{-1} = A^{-1} (B - A)B^{-1}$ we can write
\[
a(\check{\Sigma}^0) - a(\stackrel{\circ}{\Sigma}) =T^{-1/2}
\sum_{t=1}^T \left(\check{\Sigma}_t^0\right)^{-1} \left\{ \stackrel{\circ}{\Sigma}_t - \check{\Sigma}_t^0 \right\} \stackrel{\circ}{\Sigma}_t^{-1}   u_t\tilde{X}_{t-1}'.
\]
Take the norm on the right-hand side and apply Lemma \ref{lemma_A}(f,h,i), Cauchy-Schwarz inequality and the fact that  $T^{-1} \sum_{t=1}^T \| u_t\tilde{X}_{t-1}' \|^2 = O_p(1)$ by Lemma \ref{lem0}-a).

(a') Use the same decomposition to write
\begin{equation}\label{aprim}
a(\check{\Sigma}^0) - a(\check{\Sigma}) =T^{-1/2}
\sum_{t=1}^T \left(\check{\Sigma}_t^0\right)^{-1} \left\{ \check{\Sigma}_t - \check{\Sigma}_t^0 \right\} \check{\Sigma}_t^{-1}   u_t\tilde{X}_{t-1}'.
\end{equation}
Now, if $\|\cdot\|_2$ denotes the spectral norm, use the inequality
\[
\|B^{1/2} - A^{1/2}\|_2 \leq \frac{1}{2}\left[ \max\{ \|A^{-1}\|_2, \|B^{-1}\|_2  \} \right]^{1/2} \|B - A\|_2
\]
(see for instance Horn and Johnson (1994), page 557), and deduce that
\[
\| \check{\Sigma}^0_t - \check{\Sigma}_t \|_2 \leq \frac{\nu_T}{2}\left[ \max\left\{\left\|\left[\left(\check{\Sigma}_t^0\right)^{2}  +\nu_T I_d \right]^{-1}\right\|_2, \left\|\left[\left(\check{\Sigma}^0_t\right)^{2}\right]^{-1}\right\|_2 \right\} \right]^{1/2}.
\]
Now, if $r\in (0,1]$ and $A_{[Tr ]} - B = o_p(1)$ with $B$ positive definite, it is easy to check that $\|A_{[Tr ]}^{-1}\|_2 \leq \{ 1+o_p(1)\} \| B^{-1}\|_2$. Use Lemma \ref{unif_xu_phi08} and Assumption \textbf{A1'}(i) to deduce that the spectral norms of $[\left(\check{\Sigma}_t^0\right)^{2}  +\nu_T I_d ]^{-1}$ and $[\left(\check{\Sigma}_t^0\right)^{2} ]^{-1}$ are bounded in probability. Finally, take spectral norm on the right-hand side of (\ref{aprim}), use the fact that $\nu_T = o(T^{-1/2})$ and deduce (a').

(b) Consider the identity $$A^{-1} - B^{-1} = B^{-1} (B - A)B^{-1} + B^{-1} (B - A)A^{-1} (B - A) B^{-1} $$ and write
\begin{eqnarray*}
a(\stackrel{\circ}{\Sigma}) - a(\bar \Sigma) & = & T^{-1/2}\sum_{t=1}^T
[\stackrel{\circ}{\Sigma}_t^{-1} -  \bar{\Sigma}_t^{-1} ] u_t\tilde{X}_{t-1}'\\
&=&
T^{-1/2}\sum_{t=1}^T \bar{\Sigma}_t^{-1}
[ \bar{\Sigma}_t - \stackrel{\circ}{\Sigma}_t ] \bar{\Sigma}_t^{-1}u_t\tilde{X}_{t-1}'
\\
&&+ T^{-1/2}\sum_{t=1}^T \bar{\Sigma}_t^{-1}
[ \bar{\Sigma}_t - \stackrel{\circ}{\Sigma}_t ] \stackrel{\circ}{\Sigma}_t^{-1}[ \bar{\Sigma}_t - \stackrel{\circ}{\Sigma}_t ] \bar{\Sigma}_t^{-1} u_t\tilde{X}_{t-1}'
\\ &=:& T^{-1/2}\sum_{t=1}^T \Delta_{1t}(b) + T^{-1/2}\sum_{t=1}^T \Delta_{2t}(b)
\\&=:& \Delta_1(b) + \Delta_2(b).
\end{eqnarray*}
Note that by equation (22) in Xu and Phillips (2008), $$\{  \Delta_{1t}(b), \mathcal{F}_t \} = \{ \bar{\Sigma}_t^{-1}
[ \bar{\Sigma}_t - \stackrel{\circ}{\Sigma}_t ] \bar{\Sigma}_t^{-1}u_t\tilde{X}_{t-1}', \mathcal{F}_t \}$$ is a martingale difference (m.d.) sequence indexed by the bandwidths $b$.\footnote{ It is important to notice that for a fixed bandwidth the sequence $(\Delta_{1t}(b))$ is not adapted to the filtration $(\mathcal{F}_t)$. As a consequence, the expectation $E\{\Delta_{1t}(b)^\prime \Delta_{1s}(b)\}$ is not necessarily zero and therefore the equality $E\{ \| \Delta_1(b)\|^2\} = T^{-1} \sum_{t=1}^T  E\{ \| \Delta_{1t}(b) \|^2\}$ does not necessarily holds.\label{footj}

} To prove that $\Delta_1(b) = o_p(1) $ uniformly w.r.t. $b$ we show
that this uniform rate holds cellwise. For this purpose it easy to
see that it suffices to prove that
\[
S_T(h) = \frac{1}{\sqrt{T}} \sum_{t=1}^T \varepsilon_t \omega_i w_{ti}(h) = o_p(1)
\]
uniformly w.r.t. $h\in\mathcal{B}_T$ where $\{\varepsilon_t,
\mathcal{F}_t \}$  and $\{\omega_t, \mathcal{F}_t \}$  are
univariate m.d. sequence satisfying suitable moment conditions and
\begin{equation}\label{rez}
\sup_{t\geq 1} E\{ \varepsilon_t^2 +  \omega_t^2\mid \mathcal{F}_{t-1} \} <\infty.
\end{equation}
More
precisely, $ \varepsilon_t \omega_i $ could be any cell of
$$ \bar{\Sigma}_t^{-1}
[ \Sigma_i - u_i u_i^\prime ]
\bar{\Sigma}_t^{-1}u_t\tilde{X}_{t-1}'.$$ Using the Inverse Fourier
Transform and a change of variables, we rewrite
\begin{eqnarray*}\label{dec_st}
S_T(h) & = & \!\!\!\!\frac{1}{T\sqrt{T} \; h} \sum_{t,i=1}^T \varepsilon_t \omega_i \widehat f_T^{-1}(t/T;h) K((t-i)/Th) - \Delta_T(h) \\
&=& \!\!\!\!\frac{1}{T\sqrt{T} \; h} \int_{\mathbb{R}} \sum_{t,i=1}^T \!\varepsilon_t \omega_i \widehat f_T^{-1}(t/T;h) \exp\!\!\left(\!2\pi \sqrt{-1}\;\frac{t-i}{Th}\;u\!\right)\!\mathcal{F}[ K] (u) du \!- \!\Delta_T(h)\\
&\stackrel{u=sh}{=}& \!\!\!\frac{1}{\sqrt{T} } \int_{\mathbb{R}} \left\{ \frac{1}{\sqrt{T}} \sum_{t=1}^T \frac{\varepsilon_t (1+|s|)^{-\tau }}{ \widehat f_T(t/T;h)} \exp\left(2\pi \sqrt{-1}\;\frac{t}{T}\;s\right) \right\}\\
&&\!\!\! \!\!\times
\left\{ \frac{1}{\sqrt{T}} \sum_{i=1}^T \omega_i (1+|s|)^{-\tau }\exp\left(-2\pi \sqrt{-1}\;\frac{i}{T}\;s \right)\right\}
 \frac{\mathcal{F}[ K] (sh) }{(1+|s|)^{-2\tau }} ds - \Delta_T(h)\\
&=:& \!\!\!\!\frac{1}{\sqrt{T} } \int_{\mathbb{R}} \left\{ \frac{1}{\sqrt{T}} \sum_{t=1}^T \widehat S_{1t}(h,s) \right\} \left\{ \frac{1}{\sqrt{T}} \sum_{i=1}^T S_{2i}(s)\right\} \frac{\mathcal{F}[ K] (sh) }{(1+|s|)^{-2\tau }} \; ds - \Delta_T(h)
\end{eqnarray*}
where
\[
\Delta_T(h)=\frac{K(0)}{T\sqrt{T} \; h}\sum_{t=1}^T \varepsilon_t \omega_t \widehat f_T^{-1}(t/T;h),
\]
$\widehat f_T(t/T;h) = T^{-1}h^{-1}\sum_{j=1}^T K((t-j)/Th)$ and $\tau$ is any (arbitrary small) positive constant.
It is easy to see that Assumption {\bf A1}'(i) and inequality (\ref{inf_bound}) imply $$\sup_{h\in\mathcal{B}_T}| \Delta_T(h) | = o_p(T^{-1/2}b_T^{-1})= o_p(1).$$ Since $$\frac{1}{\sqrt{T}} \int_{\mathbb{R}} \frac{|\mathcal{F}[ K] (sh)|}{(1+|s|)^{-2\tau }} \; ds  \leq C \frac{1}{\sqrt{T} b_T^{1+2\tau}} \int_{\mathbb{R}} |s \mathcal{F}[ K] (s)| \; ds = O \left(\left(T b_T^{2+\gamma}\right)^{- 1/2}\right)$$ provided $\tau$ is sufficiently small, it suffices to prove
\begin{equation}\label{s1h}
\sup_{h\in\mathcal{B}_T}\sup_{s\in\mathbb{R}} \left| \frac{1}{\sqrt{T}} \sum_{t=1}^T \widehat S_{1t}(h,s) \right| = o_p(1)
\end{equation}
and
\begin{equation}\label{s1hb}
\sup_{s\in\mathbb{R}} \left| \frac{1}{\sqrt{T}} \sum_{i=1}^T S_{2i}(s)\right| = o_p(1)
\end{equation}
Let us notice that
\[
\widehat f_T(t/T;h) = h^{-1}\sum_{j=1}^T \int^{\frac{t-j+1}{T}}_{\frac{t-j}{T}}K\left( \frac{[Tr]}{Th} \right) dr
\stackrel{z = r/h}{=} \int^{\frac{t}{Th}}_{\frac{t-T}{Th}}K\left( \frac{[Tzh]}{Th} \right) dz.
\]
For $0< c_{max}^{-1} \leq \vartheta \leq c_{min}^{-1}$ define
\[
f_T(t/T; b_T/\vartheta) = \int^{\frac{t\vartheta}{Tb_T}}_{\frac{(t-T)\vartheta}{Tb_T}}K\left( z \right) dz .
\]
Then, if $K(\cdot)$ is differentiable, for any $1\leq t \leq T$ and $h\in\mathcal{B}_T$
\begin{eqnarray*}
\left| \widehat f_T(t/T;h) -  f_T(t/T;h)\right| & \leq & \int
_{-\infty}^{\infty} \left| K\left( z \right) -K\left(
\frac{[Tzh]}{Th} \right) \right|  dz\\
& \leq & \int_{-\infty}^{\infty} \int_{\frac{[Tzh]}{Th}} ^{z} \left|
K^\prime (v) \right|dv dz\\
& \leq & \int_{-\infty}^{\infty} \int_{z-\frac{1}{Th}} ^{z} \left|
K^\prime (v) \right|dv dz\\
&=&  \int_{-\infty}^{\infty} \left| K^\prime (v)
\right|\int^{v+\frac{1}{Th}}_{v} dzdv \leq \frac{C}{Tb_T}.
\end{eqnarray*}
When the $K(\cdot)$ is differentiable except a finite number of points, the same type of upper bound can be derived after minor and obvious changes. Hence, with the notation
\[
S_{1t}(\vartheta,s) = \frac{\varepsilon_t (1+|s|)^{-\tau }}{ f_T(t/T;b_T/\vartheta)} \exp\left(2\pi \sqrt{-1}\;\frac{t}{T}\;s\right), \quad 0< c_{max}^{-1}\leq \vartheta \leq c_{min}^{-1}, \;s\in\mathbb{R},
\]
since for any real numbers $a$ and $b$, $a^{-1} = b^{-1} + (b-a)/ab$ and knowing that $\widehat f_T(\cdot;b_T/\vartheta)$ and $f_T(\cdot;b_T/\vartheta)$ are uniformly bounded away from zero (see inequality (\ref{inf_bound}) below), we obtain
\[
\sup_{\vartheta}\sup_{s}\left|\frac{1}{\sqrt{T}} \sum_{t=1}^T  \! \left\{\widehat S_{1t}(\vartheta,s)\! - \!S_{1t}(b_T/\vartheta,s)\!\right\}  \right|  \leq \frac{C}{\sqrt{T}\; b_T } \frac{1}{T} \sum_{t=1}^T \!|\varepsilon_t| \!=\! O_p(T^{-1/2} b_T^{-1}) \!=\! o_p(1).
\]
Therefore is suffices to prove
\begin{equation}\label{s1h2}
\sup_{c_{max}^{-1}\leq \vartheta \leq c_{min}^{-1}}\sup_{s\in\mathbb{R}} \left| \frac{1}{\sqrt{T}} \sum_{t=1}^T  S_{1t}(\vartheta,s) \right| = o_p(1)
\end{equation}
in place of (\ref{s1h}). For proving (\ref{s1hb}) and (\ref{s1h2})
we use a uniform CLT for m.d.  indexed by a class of functions, see
Bae \emph{et al.} (2010), Bae and Choi (1999). Here our  indexing
classes functions  depend on $c_{max}^{-1}\leq \vartheta \leq c_{min}^{-1}$ and
$s\in\mathbb{R}$, respectively on $s\in\mathbb{R}$, and we prove
that their covering numbers are of polynomial order independent of
$T$. Now we can explain the unique role of the $(1+|s|)^{\tau}$
function: it cuts the high frequencies of the complex exponential
function and allows one to obtain polynomial covering numbers.
Consider the family of functions $\mathcal{F}_{11} =
\{\varphi_{11}(\cdot;\vartheta):[0,1]\rightarrow [0,1]: c_{max}^{-1}\leq \vartheta \leq c_{min}^{-1} \}$ and $
\mathcal{F}_{12} = \{ \varphi_{12}(\cdot;s):[0,1]\rightarrow
\mathbb{C}: s\in\mathbb{R}\}$ where
\[
\varphi_{11}(r;\vartheta) \!=\!\! f_T (r; b_T/\vartheta) \!=\! F_K (\vartheta r/b_T) - F_K(\vartheta (r-1)/b_T),\;\; \varphi_{12}(r;s)\!=\! \frac{\exp\!\left(2\pi \sqrt{-1}\;rs\right)}{(1+|s|)^\tau},
\]
where $F_K(\cdot)$ is the cumulative distribution function
associated to the density $K(\cdot)$. By Lemma 22-ii) and Lemma 16
of Nolan and Pollard (1987), the class $\mathcal{F}_{11}$ is a
$VC-$class (also called Euclidean) for a constant envelope. The
$VC-$property for $\mathcal{F}_{12}$ is proved for instance in Lopez
and Patilea (2010).

Now we check the conditions of Theorem 1 of Bae \emph{et al.} (2010) in order to derive (\ref{s1hb}) and (\ref{s1h2}). For simplicity, we only provide the details for (\ref{s1h2}). With the notation  of Bae \emph{et al.} (2010),  $j=t$, $n=j(n) = T$, $\mathcal{E}_{Tt} = \mathcal{F}_t$, $\mathbf{X}=\mathbb{R}\times [0,1]$, $V_{Tt}(f) = T^{-1/2}\varepsilon_t \varphi_{12}(t/T;s) \varphi_{11}(t/T;\vartheta)^{-1}$ with $(\vartheta,s)\in\mathcal{T} = [c_{max}^{-1},c_{min}^{-1}]\times\mathbb{R}$, the family $\mathcal{F}$ being composed of the functions
$f(\varepsilon, r) = \varepsilon \varphi_{12}(r;s) \varphi_{11}(r;\vartheta)^{-1}$, $(\vartheta, s)\in\mathcal{T}$, with envelope
$F(\varepsilon, r) = C\varepsilon$ for some sufficiently large constant $C$. Moreover, define
\[
\{d_{\mu_n}^{(2)} (f,g)\}^2 = d^2((\vartheta_1, s_1), (\vartheta_2,s_2)) = \int_0^1 E\left\{\varepsilon_{[Tr]}^2\left(\frac{\varphi_{12}(r;s_1)}{ \varphi_{11}(r;\vartheta_1) } - \frac{\varphi_{12}(r;s_2)}{ \varphi_{11}(r;\vartheta_2) } \right)^2 \right\} dr
\]
Notice that $\sup_{0< r\leq 1} E(\varepsilon_{[Tr]}^2)< \tilde C$ for some constant $\tilde C$, $ \varphi_{11}(\cdot;\vartheta)$ is uniformly bounded and bounded away from zero (see (\ref{inf_bound})), $ \varphi_{12}(\cdot;s)$ is uniformly bounded, and
\begin{eqnarray*}
&&\!\!\! \!\!\!\!\!\!\!\!\!\int_0^1 \left(\frac{\varphi_{12}(r;s_1)}{ \varphi_{11}(r;\vartheta_1) } - \frac{\varphi_{12}(r;s_2)}{ \varphi_{11}(r;\vartheta_2) } \right)^2 dr  \\
& & \leq C_1 (\vartheta_1 \!- \!\vartheta_2)^2\int_0^1 \!\!\left\{ K(r/c_{max}b_T)r/b_T \right\}^2 + \!\left\{ K((r\!-\!1)/c_{max}b_T)(r\!-\!1)/b_T \right\}^2 dr\\
&& \quad + C_2 \int_0^1 \left\{  \varphi_{12}(r;s_1)  -  \varphi_{12}(r;s_2)\right\}^2 dr \\
&&\leq  C_3 \{b_T(\vartheta_1 - \vartheta_2)^2 + (s_1 - s_2)^2\},
\end{eqnarray*}
$\forall (\vartheta_1,s_1),( \vartheta_2,s_2)\in\mathcal{T}$, for some constants $C_1, \cdots, C_3 > 0$. Moreover, for any $\rho > 0$ there exists $c_\rho>0$ such that
$
\int_0^1 \varphi_{12}^2(r;s)   dr \leq \rho $, $\forall s \geq c_\rho.
$
Using these properties,  on one hand we check that the pseudometric space $(\mathcal{T},d(\cdot,\cdot))$ is totally bounded and, on the other hand, we check
that condition (2) of Bae \emph{et al.} (2010) holds for some sufficiently large $L$ given that the conditional variance of $\varepsilon_t$ is deterministic and bounded. The convergence to zero for $L_n(\delta)$ in  Bae \emph{et al.} (2010) is a direct consequence of our unconditional moment conditions on $\varepsilon_t$. The uniformly integrable entropy condition is ensured by the $VC-$property satisfied by the classes $\mathcal{F}_{11}$ and $\mathcal{F}_{12}$ and the finite second order moment for $\varepsilon_t$. Now, all the required ingredients are gathered to apply Theorem 1 of Bae \emph{et al.} (2010) and to deduce our property (\ref{s1h2}).

For the uniform $o_p(1)$ rate of  $ \Delta_2$ take the norms and apply Lemma \ref{lemma_A}(c,e,f) and the moment assumptions.

The results (c) to (f) are obtained by obvious adaptation of the corresponding proofs in Xu and Phillips (2008), hence the details are omitted.

Finally, to derive the result for $\check\Omega_1$,
use again the identity $A^{-1} - B^{-1} = A^{-1} (B - A)B^{-1}$, the inequality $\|A \otimes B \| = \| A\| \|B \|$, the triangle inequality and apply Lemma \ref{lemma_A}(e,g,h,j).  $\quad\square$

\vspace{0.3 cm}

\begin{lem}\label{unif_xu_phi08}
Let $g_{kl}(r-)= \lim_{\tilde r \uparrow r} g_{kl}(\tilde r)$ and $g_{kl}(r+)= \lim_{\tilde r \downarrow r} g_{kl}(\tilde r)$, for $r\in(0,1]$ and $1\leq k,l,\leq d$. Define the $d\times d-$matrices $G(r-)=\{g_{kl}(r-) \}$ and $G(r+)=\{g_{kl}(r+) \}$ and $\Sigma (r-) = G(r-)G(r-)^\prime$, $\Sigma (r+) = G(r+)G(r+)^\prime$. Set $\Sigma(1+) = 0$. Under the assumptions of Proposition \ref{lemALS},
\[
\check{\Sigma}_{[Tr]}^0 \stackrel{P}{\longrightarrow} \Sigma(r-)\int_{-\infty}^0 K(z) dz + \Sigma(r+)\int_0^\infty K(z) dz,
\]
uniformly with respect to $r\in(0,1]$.
\end{lem}

\noindent{\bf Proof of Lemma \ref{unif_xu_phi08}} It suffices to notice that equation (19) of Xu and Phillips (2008) can be obtained uniformly w.r.t. $r\in(0,1]$ and $b_{kl}\in\mathcal{B}_T$, $1\leq k\leq l\leq d$ and to prove
\[
\sup_{b_{kl}\in\mathcal{B}_T, 1\leq k\leq l\leq d} \sup_{r\in(0,1]} \{ \|A _{[Tr]} \| + \|B _{[Tr]} \| \}= o_p(1),
\]
for $A_{[Tr]} = \check{\Sigma}_{[Tr]}^0 - \stackrel{\circ}{\Sigma}_{[Tr]} $ and $B_{[Tr]} =  \stackrel{\circ}{\Sigma}_{[Tr]} - \bar \Sigma_{[Tr]}$
where $\stackrel{\circ}{\Sigma}_t$ and $\bar \Sigma_t$ are defined in equation (\ref{sigma_rond}). The uniform convergence of $A_{[Tr]}$ and $B_{[Tr]}$ is easily obtained from Lemma \ref{lemma_A}(d,g). $\quad\square$

\vspace{0.3 cm}

In the following lemma, which is an extension of the statements (d)
to (l) in Lemma A of Xu and Phillips (2008),  we gather some results
used in the proof of Proposition \ref{lemALS}. Let $w_{ti,kl} =
w_{ti}(b_{kl})$ denote the element $kl$ of the $d\times d$ matrix
$w_{ti}$ that is a function of the bandwidth $b_{kl}$.

\begin{lem}\label{lemma_A}
Let $\|\cdot\|$ denote the Frobenius norm.
Under the assumptions of Proposition  \ref{lemALS}:

(a) For all $T\geq 1$ and $1\leq k\leq l\leq d$,
\begin{equation*}
\max_{1\leq t\leq T} \sum_{i=1}^T
\sup_{b_{kl}\in\mathcal{B}_T}w_{ti,kl} \leq C< \infty,
\end{equation*}
for some constant $C$.

(b) For all $T\geq 1$ and $1\leq k\leq l\leq d$,
 $\max_{1\leq t,i\leq T} \sup_{b_{kl}\in\mathcal{B}_T} w_{ti,kl} \leq C/Tb_T$ for some constant $C>0$.

 (c)  For all $T$,
 $\inf_{b\in\mathcal{B}_T}\min_{1\leq t\leq T}\lambda_{min} (\bar \Sigma_t) \geq C>0$ for some constant $C.$

(d)
 As $T\rightarrow\infty$,
\begin{equation}\label{rtrt_b}
\max_{1\leq t\leq T} E\left( \sup_{b\in\mathcal{B}_T}\|\stackrel{\circ}{\Sigma}_t - \bar \Sigma_t \|^4\right) = O_p(\left(1/(Tb_T)^2  \right).
\end{equation}

 (e)
 For $\delta = 1,2$, as $T\rightarrow\infty$,
\begin{equation*}
 \max_{1\leq t\leq T} \sup_{b\in\mathcal{B}_T} \|\stackrel{\circ}{\Sigma}_t - \bar \Sigma_t \|^\delta = O_p (T^{-\delta/4}b_T^{-\delta/2}).
\end{equation*}

(f)
 As $T\rightarrow\infty$, $$\left(\inf_{b\in\mathcal{B}_T}\min_{1\leq t\leq T}\lambda_{min} (\stackrel{\circ}{\Sigma}_t) \right)^{-1} = O_p(1).$$

 (g)
  As $T\rightarrow\infty$,
\begin{equation*}
 \max_{1\leq t\leq T} \sup_{b\in\mathcal{B}_T} \| \check{\Sigma}_t^0 -    \stackrel{\circ}{\Sigma}_t  \| = O_p (T^{-1/2}b_T^{-1/2}).
\end{equation*}

 (h)
 As $T\rightarrow\infty$, $$\left(\inf_{b\in\mathcal{B}_T}\min_{1\leq t\leq T}\lambda_{min} (\check{\Sigma}_t^0) \right)^{-1} = O_p(1).$$

(i) As $T\rightarrow \infty$,
$\sup_{b\in\mathcal{B}_T}\sum_{i=1}^T  \| \check{\Sigma}_t^0 -    \stackrel{\circ}{\Sigma}_t  \|^2 = O_p (T^{-2}b_T^{-2}) $

(j) As $T\rightarrow \infty$, $\sup_{b\in\mathcal{B}_T}T^{-1} \sum_{t=1}^T  \| \bar{\Sigma}_t -    \Sigma_t  \| = o(1)$.

\end{lem}

\noindent{\bf Proof of Lemma \ref{lemma_A}} (a) Using the monotonicity of $K(\cdot)$ we can write
\begin{equation}\label{kk1a}
\max_{1\leq t\leq T} \sum_{i=1}^T
\sup_{b_{kl}\in\mathcal{B}_T}w_{ti,kl} \leq  \max_{1\leq t\leq T}
\frac{\sum_{i=1}^T K((t-i)/b_T c_{max})}{\sum_{i=1}^T K((t-i)/b_T
c_{min}) -K(0)}.
\end{equation}
Now, using again the monotonicity of $K(\cdot)$ and adapting the
lines of Lemma A(c) in Xu and Phillips (2008), for any
$h\in\mathcal{B}_T$ and any $1\leq t\leq T$,
\[
\frac{1}{Th}\sum_{i=1}^T K\!\! \left( \frac{t-i}{Th} \right)\! =
\sum_{i=1}^T \int_{\frac{t-i}{Th}}^{\frac{t-i+1}{Th}} \!\! K\left(
\frac{[Thz]}{Th} \right) dz \leq \int_{-\infty}^{\infty} \!\! \max\!
\left[ \! K\left( z  \right) ,  K\!\!\left(\! z - \frac{1}{Th}
\!\right)\! \right]dz\leq 2.
\]
This allows to control the numerator on the right-hand side of
(\ref{kk1a}). On the other hand, using similar arguments and the
fact that $K(0)>0$, for any $h\in\mathcal{B}_T$, any $1\leq t\leq T$
and any $0<\gamma_1<\gamma_2<\infty$,
\begin{eqnarray}\label{inf_bound}
\frac{1}{Th}\sum_{i=1}^T K\left( \frac{t-i}{Th} \right) &\geq &
\int^{\frac{t}{Th}}_{\frac{t-T}{Th}} \min \left[ \! K\left( z
\right) ,  K\!\!\left(\! z - \frac{1}{Th}
\!\right)\! \right]dz \notag \\
&\geq & \min \left[  \int_{-\gamma_2}^{-\gamma_1} K(z)dz ,  \int_{\gamma_1}^{\gamma_2} K(z)dz \right],
\end{eqnarray}
provided that $T$ is sufficiently large. The last two integrals in
the  minimum are strictly positive for a suitable choice of
$\gamma_1, \gamma_2$. This fixed lower bound considered for
$h=c_{min}b_T$ allows to control the denominator on the right-hand
side of  (\ref{kk1a}) and thus to prove (a) with a constant
depending on $\gamma_1,\gamma_2$ and $c_{max}/c_{min}$.

(b) For all $1\leq k\leq l\leq d$,
\[ w_{ti,kl} \leq \frac{\frac{1}{Tb_T
c_{min}} K\left( \frac{t-i}{Tb_T c_{max}} \right)}{\frac{1}{Tb_T
c_{max}}\sum_{j=1}^T K\left( \frac{t-j}{Tb_T c_{min}} \right)}.
\]
Now, use the fact that $K$ is bounded, $c_{max}/c_{min} < \infty$
and Lemma A(c) of Xu and Phillips (2008) to derive the upper bound.

(c) This is an easy consequence of Assumption \textbf{A1'}(i) and the proof of Lemma \ref{unif_xu_phi08}, equation (19), that holds uniformly w.r.t. $r\in(0,1]$ and $b_{kl}\in\mathcal{B}_T$, $1\leq k\leq l\leq d$.

(d)
Let $a_i(k,l)$ denote a generic element of the $d\times d-$matrix ${u}_i{u}_i' - \Sigma_i$. Then we can write
\begin{eqnarray*}
E\left(  \sup_{b\in\mathcal{B}_T} \left\| \sum_{i=1}^T w_{ti} ( {u}_i {u}_i' - \Sigma_i  ) \right\|^4\right) &\leq & E\left( \sup_{b\in\mathcal{B}_T}\left[ \sum_{k,l=1}^d \left| \sum_{i=1}^T w_{ti} a_i(k,l) \right|  \right]^4 \right)\\
&\leq & c \sum_{k,l=1}^d E \left(\sup_{b\in\mathcal{B}_T}\left| \sum_{i=1}^T w_{ti,kl} a_i(k,l) \right|^4\right)
\\
&\leq & c \sum_{k,l=1}^d E \left(\left| \sum_{i=1}^T \sup_{b\in\mathcal{B}_T}w_{ti,kl} |a_i(k,l) | \right|^4\right)
\end{eqnarray*}
where $c$ depends only on $d$. Now, by Lemma A(f) of Xu and Phillips (2008) and (a)-(b) above, for $1\leq k\leq l\leq d$
\[
E \!\left( \!\sum_{i=1}^T \sup_{b\in\mathcal{B}_T}\! w_{ti,kl} |a_i(k,l)| \!\right)^{\!\!4} \!\!\!\leq \!\!\left[ \!\max_{1\leq t\leq T} \sum_{i=1}^T \!\sup_{b\in\mathcal{B}_T}\!w_{ti,kl} \!\right]^2  \!\sum_{i=1}^T \sup_{b\in\mathcal{B}_T} \!w_{ti,kl} E|a_i(k,l)|^4 \!\leq  \!\frac{c}{(Tb_T)^2},
\]
for $c$ a constant depending only on $K$, $c_{max}/c_{min}$ and the upper bounds of the 4th order moments of the components of ${u}_i{u}_i' - \Sigma_i$. Now, (\ref{rtrt_b}) follows.

(e)
By Markov's inequality and obvious algebra we can write
\begin{eqnarray*}
&& \hspace{-1cm}P\left( \max_{1\leq t\leq T} \sup_{b\in\mathcal{B}_T}  \|\stackrel{\circ}{\Sigma}_t - \bar \Sigma_t \|^\delta  > CT^{-\delta/4}b_T^{-\delta/2}   \right) \\
&&= P\left( \max_{1\leq t\leq T} \sup_{b\in\mathcal{B}_T}  \|\stackrel{\circ}{\Sigma}_t - \bar \Sigma_t \|^4  > C^{4/\delta} T^{-1}b_T^{-2}   \right) \\
&&\leq C^{-4/\delta} Tb_T^{2} E \left( \max_{1\leq t\leq T} \sup_{b\in\mathcal{B}_T}  \|\stackrel{\circ}{\Sigma}_t - \bar \Sigma_t \|^4    \right)
\\
&& \leq C^{-4/\delta} Tb_T^{2} \sum_{t=1}^TE \left( \sup_{b\in\mathcal{B}_T}  \|\stackrel{\circ}{\Sigma}_t - \bar \Sigma_t \|^4    \right)\\
&&= C^{-4/\delta} Tb_T^{2}  T  \max_{1\leq t\leq T} E \left( \sup_{b\in\mathcal{B}_T}  \|\stackrel{\circ}{\Sigma}_t - \bar \Sigma_t \|^4    \right)\\
&&= C^{-4/\delta} O(1)
\end{eqnarray*}
where for the last equality we use (d).

(f)+(h) Using equation (3.5.33) in Horn and Johnson (1994) and (e) above we have
\begin{eqnarray*}
\min_{1\leq t\leq T}\lambda_{min} (\stackrel{\circ}{\Sigma}_t) &\geq &
\min_{1\leq t\leq T}\lambda_{min} (\bar{\Sigma}_t)
- \max_{1\leq t\leq T}\left| \lambda_{min} (\stackrel{\circ}{\Sigma}_t) - \lambda_{min} (\bar{\Sigma}_t)\right|\\
&\geq & \min_{1\leq t\leq T}\lambda_{min} (\bar{\Sigma}_t)
-\sup_{b\in\mathcal{B}_T} \max_{1\leq t\leq T} \| \stackrel{\circ}{\Sigma}_t - \bar{\Sigma}_t  \|\\
& = & \min_{1\leq t\leq T}\lambda_{min} (\bar{\Sigma}_t) + o_p(1),
\end{eqnarray*}
and hence (f) follows from (c). Similar algebra applies for (h) which will follow as a consequence of (g).

(g)+(i) Adapt the proof of Lemma A(i) and A(k) of Xu and Phillips (2008) using a decomposition like in our equation (\ref{time}).

(j) Apply Lemma A(l) of Xu and Phillips (2008) componentwise, that is $d^2$ times. $\quad\square$

\vspace{0.3 cm}

\noindent{\bf Proof of (\ref{ALS3}) and (\ref{asympshatols1}).}\quad
%
Let us denote
$$\hat{\Xi}=R\hat{\Lambda}_3^{-1}\hat{\Lambda}_2\hat{\Lambda}_3^{-1}R'\quad
\text{and}
\quad \hat{\Xi}_{\delta}=R\hat{\Lambda}_{3\delta}^{-1}\hat{\Lambda}_{2\delta}\hat{\Lambda}_{3\delta}^{-1}R'.$$
From the expressions of $Q_{OLS}$ and $Q_{OLS}^{\delta}$ we write
$$\mid Q_{OLS}-Q_{OLS}^{\delta}\mid\leq T\parallel R\hat{\theta}_{OLS}\parallel^2\parallel\hat{\Xi}^{-1}-\hat{\Xi}_{\delta}^{-1}\parallel=o_p(1),$$
since $T\parallel R\hat{\theta}_{OLS}\parallel^2=O_p(1)$ and $\parallel\hat{\Xi}^{-1}-\hat{\Xi}_{\delta}^{-1}\parallel=o_p(1)$ from the consistency of the estimators of $\Lambda_2$ and $\Lambda_3$. In addition we write
$$\frac{Q_{OLS}+Q_{OLS}^{\delta}}{2}=\frac{T}{2}\hat{\theta}_{OLS}'R'\{\hat{\Xi}^{-1}+\hat{\Xi}_{\delta}^{-1}\}R\hat{\theta}_{OLS}.$$
Noting that $\{\hat{\Xi}^{-1}+\hat{\Xi}_{\delta}^{-1}\}/2=R\Lambda_3^{-1}\Lambda_2\Lambda_3^{-1}R'+o_p(1),$ we have
$\{Q_{OLS}+Q_{OLS}^{\delta}\}/2\Rightarrow\chi_{pd_1d_2}^2$.
Since
$\max(a,b)=\{a+b+\mid a-b\mid\}/2,$
the result (\ref{asympshatols1}) follows from (\ref{asympshatols21}) and (\ref{asympshatols0}). The result (\ref{ALS3}) can be obtain in a similar way. $\quad\square$

\vspace{1 cm}

\section*{References}
\begin{description}

\item[] {\sc Andrews, D.W.K.} (2008). Laws of large numbers for dependent nonidentically distributed random variables. \textit{Econometric Theory} 4, 458-467.

\item[]{\sc Bae, J., and Choi, M.J.} (1999). The uniform CLT for martingale difference of function-indexed process
under uniform integrable entropy.
\textit{Communications Korean Mathematical Society} 14, 581-595.

\item[] {\sc Bae, J., Jun, D., and Levental, S. (2010)}. The uniform CLT for martingale differences arrays under the uniformly integrable entropy.
\textit{Bulletin of the Korean Mathematical  Society} 47, 39-51.

\item[]{\sc Bataa, E., Osborn, D.R., Sensier, M., and van Dijk, D.} (2009). Structural breaks in the international transmission of inflation. Discussion Paper, Centre for Growth and Business Cycle Research Economic Studies, University of Manchester.

\item[]  {\sc Batbekh, S., Osborn, D.R., Sensier, M., and van Dick, D.} (2007). Is there causality in the mean and volatility of inflation between the US and other G7 countries? Centre for Growth and Business Cycle Research Economic Studies, University of Manchester.
\item[]{\sc Beare B.K.} (2008). Unit root testing with unstable volatility. Working paper, Nuffield college, University of Oxford.
\item[]{\sc Blanchard, O., and Simon, J.} (2001). The long and large decline in U.S. output volatility. Brookings Papers on Economic Activity 1, 135-164.
\item[]{\sc Boswijk, H.P., and Zu, Y.} (2007). Testing for cointegration with nonstationary volatility. Working Paper, University of Amsterdam.
\item[]{\sc Busetti, F., and Taylor, A.M.R.} (2003). Variance shifts, structural breaks, and stationarity tests. \textit{Journal of Business and Economic Statistics} 21, 510-531.
\item[]{\sc Cavaliere, G.} (2004). Unit root tests under time-varying
variance shifts. \textit{Econometric Reviews} 23, 259-292.
\item[]{\sc Cavaliere, G., Rahbek, A., and Taylor, A.M.R.} (2010).
Testing for co-integra\-tion in vector autoregressions with
non-stationary volatility. \textit{Journal of Econometrics}, forthcoming.
\item[]{\sc Cavaliere, G., and Taylor, A.M.R.} (2007). Testing for unit roots in time
series models with non-stationary volatility. \textit{Journal of Econometrics} 140, 919-947.
\item[]{\sc Cavaliere, G., and Taylor, A.M.R.} (2008). Time-transformed unit root tests for models with non-stationary volatility. \textit{Journal of Time Series Analysis} 29, 300-330.
\item[]{\sc Davidson, J.} (1994). {\em Stochastic Limit Theory.} Oxford
University Press.
\item[]{\sc Doyle, B.M., and Faust, J.} (2005). Breaks in the variability and comovement of G-7 economic growth. \textit{The Review of Economics and Statistics} 87, 721-740.
\item[]{\sc Engle, R.F.} (1982). Autoregressive conditional heteroscedasticity with estimates of the variance of UK inflation. \textit{Econometrica} 50, 987-1008.
\item[]{\sc Feige, E.L., and Pierce, D.K.} (1979). The casual causal relationship between money and income: some caveats for time series analysis. \textit{Review of Economics and Statistics} 61, 21-33.
\item[]{\sc Granger, C.W.J.} (1969). Investigating causal relations by econometric models and cross-spectral methods. \textit{Econometrica} 12, 424-438.
\item[]{\sc Hafner, C.M., and Herwartz, H.} (2009).
Testing for linear vector autoregressive dynamics under multivariate generalized autoregressive heteroskedasticity.
\textit{Statistica Neerlandica} 63, 294-323.
\item[]{\sc Hansen, B.E.} (1995). Regression with nonstationary volatility. \textit{Econometrica} 63, 1113-1132.
\item[]{\sc Horn, R.A., and Johnson, C.R.} (1994). \emph{Topics in Matrix Analysis}. Cambridge University Press, Cambridge.
\item[]{\sc Kim, T-H., Leybourne, S., and Newbold, P.} (2002). Unit root tests with a break in innovation variance. \textit{Journal of Econometrics} 103, 365-387.
\item[]{\sc Kim, C.J., and Nelson, C.R.} (1999). Has the U.S. economy become more stable? A bayesian approach based on a Markov-switching model of the business cycle. \textit{The Review of Economics and Statistics} 81, 608-616.
\item[]{\sc     Lavergne, P.} (2008). A Cauchy-Schwartz inequality for expectation of matrices. Working Paper, Simon Fraser University.
\item[]{\sc Levental, S.} (1989). A uniform CLT for uniformly bounded families of martingale differences. \textit{Journal of Theoretical Probability} 2(3), 271-287.
\item[] {\sc Lopez, O., and Patilea, V.} (2010). Testing conditional moment restrictions in the presence of right-censoring depending on the covariates. Discussion Paper, CREST-Ensai.
\item[]{\sc Lütkepohl, H.} (2005). \!\textit{New Introduction to Multiple Time Series Analysis}. Springer, Berlin.
\item[]{\sc McConnell, M.M., and Perez-Quiros, G.} (2000). Output fluctuations in the United States: what has changed since the early 1980's? \textit{ The American Economic Review} 90, 1464-1476.
\item[]{\sc Nolan, D., and Pollard, D.} (1987). $U$-Processes: Rates of Convergence. \textit{The Annals of Statistics} 15(2), 780-799.
\item[] {\sc Patilea, V., and Ra\"{i}ssi, H.} (2010). Portmanteau tests for stable multivariate autoregressive processes. Working paper IRMAR-INSA.
\item[]{\sc Phillips, P.C.B., and Xu, K.L.} (2005).
Inference in autoregression under heteroskedasticity.
\textit{Journal of Time Series Analysis} 27, 289-308.
\item[]{\sc Ramey, V.A., and Vine, D.J.} (2006). Declining volatility in the U.S. automobile industry. \textit{The American Economic Review} 96, 1876-1889.
\item[]{\sc Robinson, P.M.} (1987). Asymptotically efficient estimation in the presence of heteroskedasticity of unknown form. \textit{Econometrica} 55, 875-891.
\item[]{\sc Sanso, A., Arag\'{o}, V., and Carrion, J.L. } (2004). Testing for changes in the unconditional variance of financial time series. DEA Working Paper, Universitat de les Illes Balears.
\item[]{\sc Sensier, M., and van Dijk, D.} (2004). Testing for volatility changes in U.S. macroeconomic time series. \textit{Review of Economics and Statistics} 86, 833-839.
\item[]{\sc
Sims, C.A.} (1972). Money income and causality. \textit{American
Economic Review} 62, 540-542.
\item[]{\sc St\u{a}ric\u{a}, C., and Granger C.W.J.} (2005). Nonstationarities in stock returns. \textit{Reviews of Economics and Statistics} 87, 503-522.
\item[]{\sc
Stock, J.H., and Watson, M.W.} (1989). Interpreting the evidence on
money-income causality. \emph{Journal of Econometrics} 40, 161-181.
\item[]{\sc
Thornton, D.L., and Batten, D.S.} (1985). Lag-length selection and tests of Granger causality between money and income. \emph{Journal of Money, Credit, and Banking} 17, 164-178.
\item[]{\sc Warnock, M.V.C., and Warnock, F.E.} (2000). The declining volatility of U.S. employment: was Arthur Burns right? Board of Governors of the Federal Reserve System, International finance discussion papers, 677.
\item[]{\sc Xu, K.L., and Phillips, P.C.B.} (2008).
Adaptive estimation of autoregressive models with time-varying
variances. \textit{Journal of Econometrics} 142, 265-280.
\end{description}

\newpage
\section*{Tables and Figures}
\vspace*{1cm}

\begin{table}[hh]\!\!\!\!\!\!\!\!\!\!
\begin{center}
\caption{\small{Empirical size (in \%) of the different Wald tests in the case of iid homoscedastic innovations. We take $a_{11}=a_{22}=0.2$, $a_{21}=0.1$ and $a_{12}=0$. The errors are standard Gaussian.}}
\begin{tabular}{|c|c|c|c|c|}
\hline
  T & 50 & 100 & 200&400 \\
  \hline\hline
  $W_{OLS}$ & {\bf 8.3} & 6.1 & 5.5& 5.4\\
  $W_{OLS}^{\delta}$ &{\bf 9.3} & 6.2 & 5.3& 5.4\\
  $W_{OLS}^{\max}$ & {\bf 9.5} & {\bf 6.6} & 5.6&5.6\\
  \hline\hline
  $W_{S}$ & {\bf 7.1} & 5.3 & 4.9& 4.9\\
  \hline\hline
  $W_{ALS}$ &  {\bf 12.4} & 5.2 & 5.3& 5.1\\
  $W_{ALS}^{\delta}$ & {\bf 13.3} & 5.5 & 5.4& 5.1\\
  $W_{ALS}^{\max}$ & {\bf 13.5} & 5.5 & 5.4& 5.1\\
  \hline\hline
  $W_{GLS}$ &  6.2 & 4.9 &  5.0& 4.5\\
  $W_{GLS}^{\delta}$ & {\bf 7.6} & 5.6 & 5.4& 4.9\\
  $W_{GLS}^{\max}$ & {\bf 8.0} & 6.0 & 5.4& 5.1\\
  \hline
\end{tabular}
\label{tab1}
\end{center}
\end{table}

\begin{table}[hh]\!\!\!\!\!\!\!\!\!\!
\begin{center}
\caption{\small{Empirical size (in \%) of the different Wald tests. The innovations are
heteroscedastic with $\gamma_1=20$, $\rho=0.6$, and we take $a_{11}=a_{22}=0.2$, $a_{21}=0.1$, $a_{12}=0$.}}
\begin{tabular}{|c|c|c|c|c|}
\hline
  T & 50 & 100 & 200&400   \\
  \hline\hline
  $W_{OLS}$ & {\bf 8.8} & 5.8 & 4.8&5.2  \\
  $W_{OLS}^{\delta}$ &{\bf 9.7} & {\bf 6.5} & 5.0&5.4 \\
  $W_{OLS}^{\max}$ & {\bf 10.2} & {\bf 6.8} & 5.0& 5.5 \\
  \hline\hline
  $W_{S}$ & {\bf 9.3} & {\bf 8.1} &{\bf 6.6} &{\bf 8.0}  \\
  \hline\hline
  $W_{ALS}$ & {\bf 7.1} & 5.5 & 4.9 &4.8 \\
  $W_{ALS}^{\delta}$ & {\bf 8.3} & 6.2 & 5.6 &5.4 \\
  $W_{ALS}^{\max}$ & {\bf 8.3} & 6.3 & 5.6 &5.4 \\
  \hline\hline
  $W_{GLS}$ & 5.2& 4.1 &  5.2 & 4.2 \\
  $W_{GLS}^{\delta}$ & 5.7 & 4.0 & 4.2&{\bf 3.4}  \\
  $W_{GLS}^{\max}$ & {\bf 6.3} & 4.4 & 5.4 &4.2 \\
  \hline
\end{tabular}
\label{tab2}
\end{center}
\end{table}

\begin{table}[hh]\!\!\!\!\!\!\!\!\!\!
\begin{center}
\caption{\small{Empirical power (in \%) of the different Wald tests. The innovations are iid
homoscedastic. We take $a_{11}=a_{22}=0.2$, $a_{21}=0.1$.}}
{\small \begin{tabular}{|c|c|c|c|c|c|c|c|c|c|}
\hline
  $a_{12}$ & -0.4 & -0.3 & -0.2 & -0.1& 0.1 & 0.2 & 0.3 & 0.4\\
  \hline\hline
$W_{OLS}$ & 98.3& 86.5& 53.3& 20.5& 18.1& 54.3& 84.5& 96.8 \\
$W_{OLS}^{\delta}$ & 98.3& 87.1& 54.2& 21.4& 18.7& 56.1& 84.9& 97.0 \\
$W_{OLS}^{\max}$ & 98.3& 87.5& 54.7& 21.9& 19.2& 56.5& 85.3& 97.0 \\
\hline\hline
$W_{S}$ & 98.1& 86.0& 53.2& 18.9& 17.0& 54.0& 84.4& 96.6\\
\hline\hline
$W_{ALS}$ & 98.3& 86.6& 53.0& 19.8& 17.4& 55.7& 84.4& 97.0\\
$W_{ALS}^{\delta}$ & 98.3& 86.6& 53.7& 20.0& 17.8& 56.0& 84.9& 97.0\\
$W_{ALS}^{\max}$ & 98.3& 86.7& 53.8& 20.2& 17.9& 56.2& 84.9& 97.0\\
\hline\hline
$W_{GLS}$ & 97.8& 85.6& 51.0& 18.1& 15.7& 52.4& 84.4& 96.9\\
$W_{GLS}^{\delta}$ & 98.7& 88.1& 53.0& 19.6& 17.5& 55.2& 85.8& 97.6\\
$W_{GLS}^{\max}$ & 98.7& 88.3& 54.2& 20.1& 18.3& 56.4& 86.4& 97.6\\
\hline
\end{tabular}}
\label{tabpow0}
\end{center}
\end{table}

\vspace*{1.7cm}
\begin{table}[hh]\!\!\!\!\!\!\!\!\!\!
\begin{center}
\caption{\small{Empirical power (in \%) of the different Wald tests. The innovations are
heteroscedastic with $\gamma_{1}=20$ and $\rho=0.6$. We take $a_{11}=a_{22}=0.2$ and $a_{21}=0.1$.}}
{\small \begin{tabular}{|c|c|c|c|c|c|c|c|c|c|}
\hline
  $a_{12}$ & -0.8& -0.6& -0.4& -0.2&  0.2 & 0.4 & 0.6 & 0.8\\
  \hline\hline
$W_{OLS}$ & 96.9& 81.4& 48.1& 17.3 &  14.2&  40.3& 70.0& 90.6 \\
$W_{OLS}^{\delta}$ &97.5& 82.4& 49.9& 18.2 &  15.5&  41.5& 70.8& 90.9 \\
$W_{OLS}^{\max}$ & 97.7& 83.4& 51.5& 18.7&   15.7&  42.2& 71.4& 91.3 \\
\hline\hline
$W_{S}$ & 98.4& 85.4& 53.9& 20.1 &  17.1 & 46.3& 75.5& 92.8\\
\hline\hline
$W_{ALS}$ &98.8& 86.7& 50.8& 17.7  & 13.5 & 45.2& 75.4& 93.0\\
$W_{ALS}^{\delta}$ &98.9& 87.8& 54.2& 19.4  & 15.6&  48.4& 77.4& 94.5\\
$W_{ALS}^{\max}$ &98.9& 87.8& 54.2& 19.4&   15.6 & 48.4 &77.4& 94.5\\
\hline\hline
$W_{GLS}$ &99.1& 88.7& 52.6& 17.7&   14.2&  48.0& 79.5& 96.1\\
$W_{GLS}^{\delta}$ &99.1& 89.9& 52.9& 18.1  & 12.4&  46.9& 78.3& 95.1\\
$W_{GLS}^{\max}$ & 99.2& 90.4& 55.5& 19.0&   14.5&  49.6& 80.5& 96.1\\
\hline
\end{tabular}}
\label{tabpow1}
\end{center}
\end{table}


\begin{table}[hh]\!\!\!\!\!\!\!\!\!\!
\begin{center}
\caption{\small{The estimators of the autoregressive parameters of the VAR(1) model for the balance data for the U.S..}}
\begin{tabular}{|c|c|c|c|c|}
\hline
    Parameter & $\theta_1$  & $\theta_2$ & $\theta_3$ & $\theta_4$ \\
\hline\hline
   ALS estimation & $0.33_{[0.08]}$ & $0.02_{[0.02]}$&$-0.35_{[0.30]}$ & $-0.07_{[0.08]}$\\
   OLS estimation &$0.45_{[0.23]}$ & $0.00_{[0.02]}$& $-1.02_{[0.60]}$& $0.1_{[0.17]}$\\
   Standard estimation &$0.45_{[0.07]}$ & $0.00_{[0.02]}$& $-1.02_{[0.37]}$& $0.1_{[0.08]}$\\
  \hline
\end{tabular}
\label{estimates}
\end{center}
\end{table}

\begin{table}[hh]\!\!\!\!\!\!\!\!\!\!
\begin{center}
\caption{\small{The balance data for the U.S.: the p-values of the ARCH-LM tests (in \%) for the components of the ALS-residuals of a VAR(1).}}
\begin{tabular}{|c|c|c|c|}
\hline
    lags & 2  & 5 & 10\\
\hline\hline
   $\check{\epsilon}_{1t}$ & 22.26 & 45.05& 36.44\\
   \hline
   $\check{\epsilon}_{2t}$&25.32 & 73.32& 77.18\\
  \hline
\end{tabular}
\label{archLM}
\end{center}
\end{table}

\begin{table}[hh]\!\!\!\!\!\!\!\!\!\!
\begin{center}
\caption{\small{The p-values of the portmanteau tests (in \%) for the checking of the adequacy of the VAR(1) model for the U.S. trade balance data.}}
\begin{tabular}{|c|c|c|}
\hline
    $m$ & 5  & 15 \\
\hline\hline
   $LB_{m}^{S}$&0.00 & 0.01\\
  \hline
   $LB_{m}^{OLS}$& 50.80 & 99.94\\
  \hline
   $LB_{m}^{ALS}$ & 6.36 & 15.95\\
  \hline
\end{tabular}
\label{pvalues}
\end{center}
\end{table}

\begin{table}[hh]\!\!\!\!\!\!\!\!\!\!
\begin{center}
\caption{\small{The p-values of the Wald tests for Granger causality in mean (in \%) from the U.S. balance on services to the U.S. balance on merchandise.}}
\begin{tabular}{|c|c|}
  \hline
   $W_{OLS}$& 8.74 \\
  \hline
  $W_{S}$& 0.57  \\
  \hline
   $W_{ALS}$ & 25.20 \\
  \hline
\end{tabular}
\label{pvaluesgr}
\end{center}
\end{table}

\begin{table}[hh]\!\!\!\!\!\!\!\!\!\!
\begin{center}
\caption{\small{The statistics of the Wald tests for Granger causality in mean from the U.S. balance on services to the U.S. balance on merchandise.}}
\begin{tabular}{|c|c|}
  \hline
   $Q_{OLS}$& 2.92 \\
  \hline
  $Q_{S}$& 7.64  \\
  \hline
   $Q_{ALS}$ & 1.31 \\
  \hline
\end{tabular}
\label{statgr}
\end{center}
\end{table}

%
%

\clearpage

\begin{figure}[h]\!\!\!\!\!\!\!\!\!\!
\vspace*{6.9cm} 


\protect \includegraphics{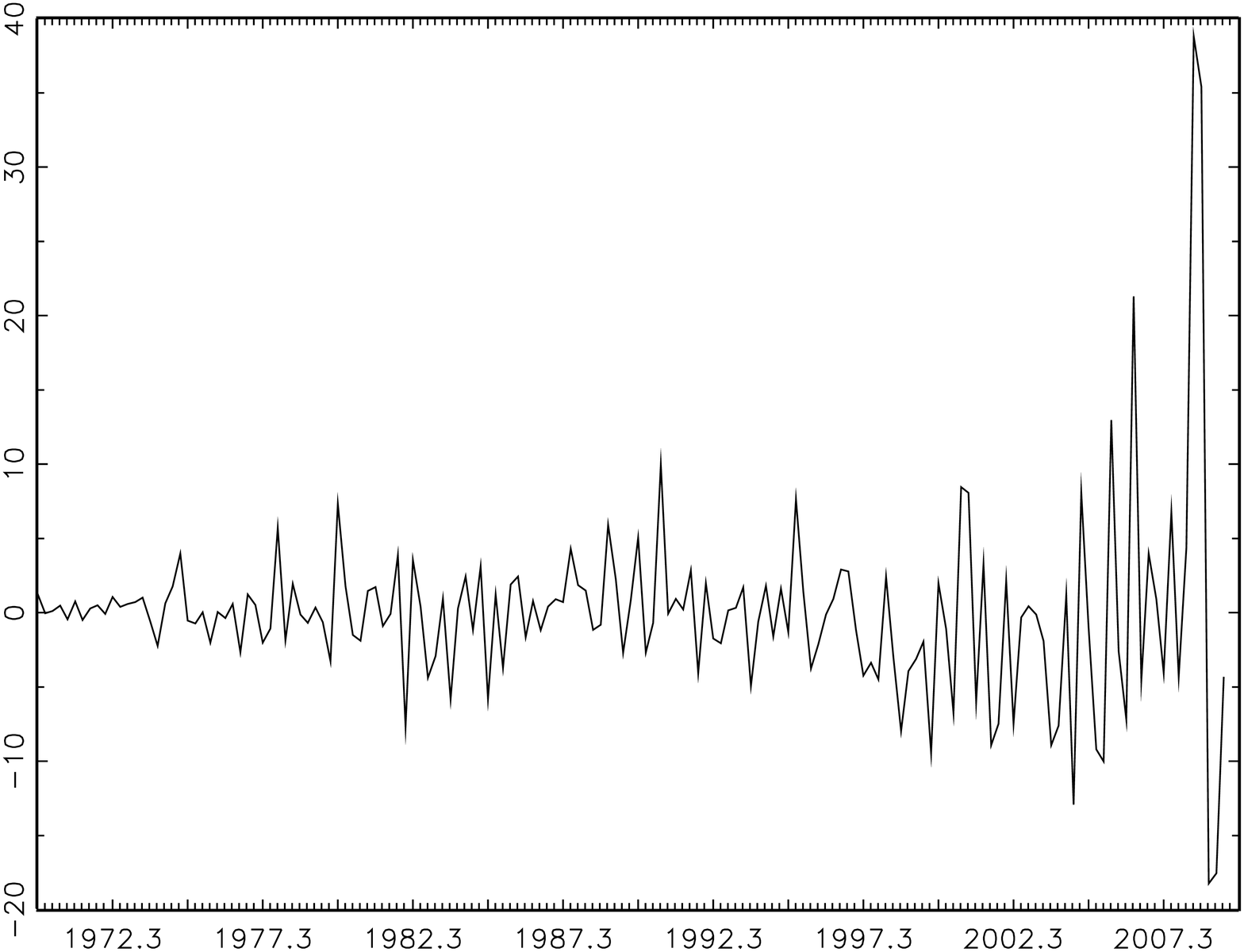} \protect \includegraphics{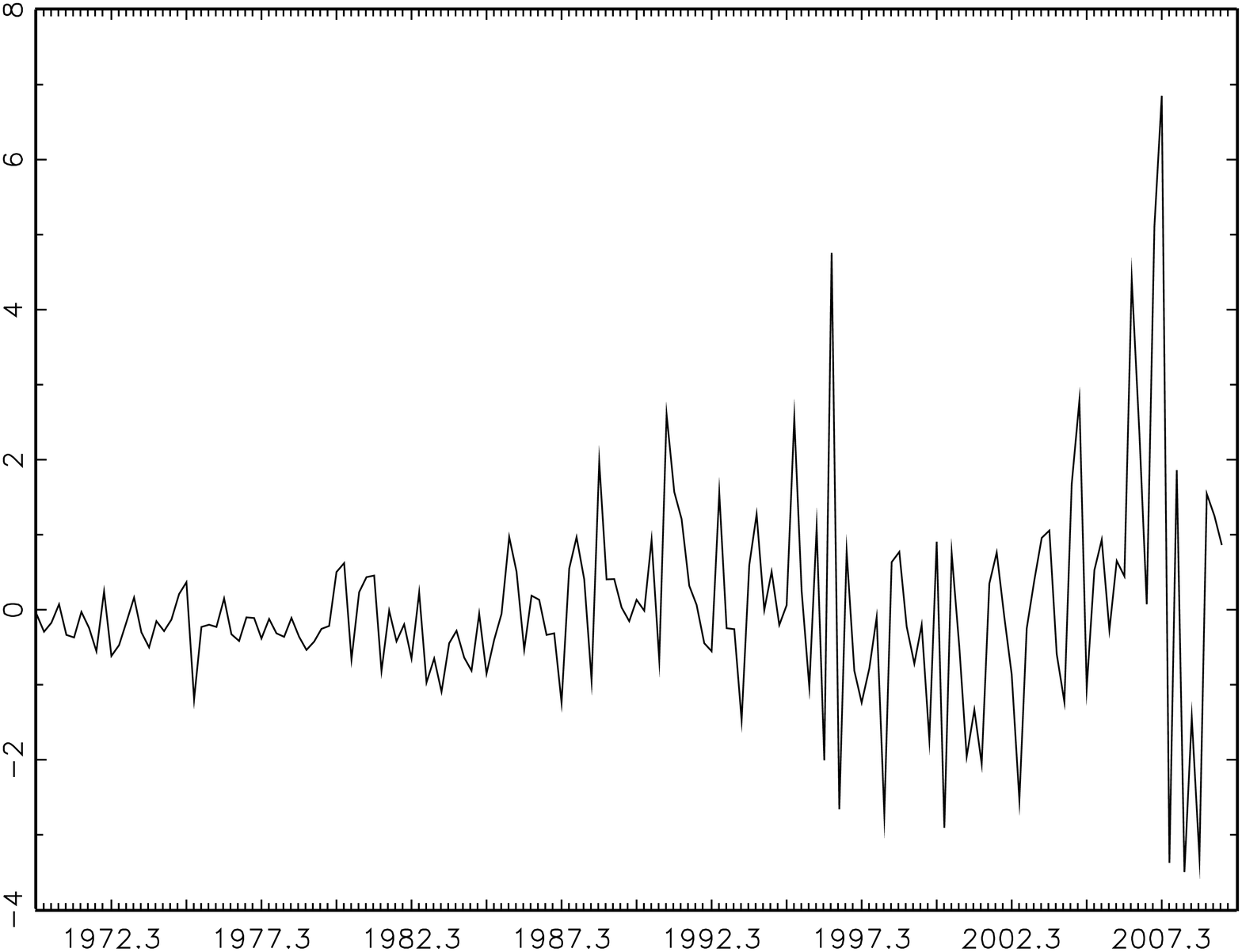} \caption{\label{residols}
{\footnotesize The same as in Figure \ref{residals} but for the OLS residuals.}}
\end{figure}
\end{document}